\documentclass[12pt,draftclsnofoot,onecolumn]{IEEEtran}
\usepackage{amsthm}
\usepackage[cmex10]{amsmath}
\usepackage{graphicx}
\usepackage[tight,footnotesize]{subfigure}
\usepackage{amsfonts}
\usepackage{algorithmic}
\usepackage{algorithm}
\usepackage{graphicx}
\usepackage{subfigure}
\usepackage{bm}
\usepackage{times,color,amssymb,epsfig,psfrag,stfloats,enumerate}
\usepackage{ulem}

\theoremstyle{theorem} \newtheorem{theorem}{Theorem}
\theoremstyle{proposition} \newtheorem{proposition}{Proposition}
\theoremstyle{lemma} \newtheorem{lemma}{Lemma}
\theoremstyle{corollary} \newtheorem{corollary}{Corollary}

\ifCLASSINFOpdf
\else
\fi

\begin{document}
\title{Networked MIMO with Fractional Joint Transmission in Energy Harvesting Systems}

\author{Jie~Gong~\IEEEmembership{Member,~IEEE}, Sheng~Zhou~\IEEEmembership{Member,~IEEE}, Zhenyu~Zhou~\IEEEmembership{Member,~IEEE}

\thanks{Jie Gong is with School of Data and Computer Science, Sun Yat-sen University, Guangdong 510006, China. Email: gongj26@mail.sysu.edu.cn}
\thanks{Sheng Zhou is with Tsinghua National Laboratory for Information Science and Technology, Department of Electronic Engineering, Tsinghua University, Beijing 100084, China. Email: sheng.zhou@tsinghua.edu.cn.}
\thanks{Zhenyu Zhou is with State Key Laboratory of Alternate Electrical Power System with Renewable Energy Sources, School of Electrical and Electronic Engineering, North China Electric Power University, Beijing 102206, China. Email: zhenyu\_zhou@ncepu.edu.cn}
%\thanks{This work is sponsored in part by the National Basic Research Program of China (No.~2012CB316001), and the Nature Science Foundation of China (No.~61201191 and 61401250), the Creative Research Groups of NSFC (No.~61321061), the Sino-Finnish Joint Research Program of NSFC (No.~61461136004), and Hitachi R\&D Headquarter.}
}


\maketitle

\begin{abstract}
This paper considers two base stations (BSs) powered by renewable energy serving two users cooperatively. With different BS energy arrival rates, a \emph{fractional joint transmission} (JT) strategy is proposed, which divides each transmission frame into two subframes. In the first subframe, one BS keeps silent to store energy while the other transmits data, and then they perform zero-forcing JT (ZF-JT) in the second subframe. We consider the average sum-rate maximization problem by optimizing the energy allocation and the time fraction of ZF-JT in two steps. Firstly, the sum-rate maximization for given energy budget in each frame is analyzed. We prove that the optimal transmit power can be derived in closed-form, and the optimal time fraction can be found via bi-section search. Secondly, \emph{approximate dynamic programming} (DP) algorithm is introduced to determine the energy allocation among frames. We adopt a linear approximation with the features associated with system states, and determine the weights of features by simulation. We also operate the approximation several times with random initial policy, named as \emph{policy exploration}, to broaden the policy search range. Numerical results show that the proposed fractional JT greatly improves the performance. Also, appropriate policy exploration is shown to perform close to the optimal.
\end{abstract}


% \fi
\IEEEpeerreviewmaketitle

\section{Introduction}
Wireless communication with energy harvesting technology, which exploits renewable energy to power wireless devices, is expected as one of the promising trends to meet the target of green communications in the future. The advantages of energy harvesting include the sustainability with renewable energy source, the flexibility of network deployment without power line to reduce network planning cost, and etc. Recently, wireless cellular networks with renewable energy are rapidly developing. For instance, China Mobile has built about 12,000 renewable energy powered base stations (BSs) by 2014 \cite{ChinaMobile}. However, due to the randomness of the arrival process of the renewable energy and the limitation on the battery capacity, energy shortage or waste will occur when the energy arrival mismatches with the network traffic requirement. How to efficiently use the harvested energy is a big challenge.

In the literature, a lot of research work has focused on the energy harvesting based communications. For single-link case, the optimal power allocation structure, \emph{directional water-filling}, is found in both single-antenna transceiver system \cite{ozel2011transmission, jie2013optimal} and multiple-input multiple-output (MIMO) channel \cite{hu2015optimal}. The research efforts have been further extended to the network case, and the power allocation policies are proposed for broadcast channel \cite{yang2012broadcasting}, multiple access channel \cite{yang2012optimal}, interference channel \cite{tutuncuoglu2012sum}, as well as cooperative relay networks \cite{chuan2013threshold, minasian2014energy}. Nevertheless, there lacks research effort on the effect of energy harvesting on the multi-node cooperation, i.e., network MIMO.

The network MIMO technology, which shares the user data and channel state information among multiple BSs, and coordinates the data transmission and reception by transforming the inter-cell interference into useful signals, has been extensively studied in the literature \cite{karakayali2006network, zhang2009networked, huang2009increasing}. And it has already been standardized in 3GPP as Coordinated Multi-Point (CoMP) \cite{3GPP2011TR}. By applying joint precoding schemes such as zero-forcing (ZF) \cite{boccardi2006zero, kaviani2011optimal} among BSs for joint transmission (JT), the system sum-rate can be greatly increased. However, how the dynamic energy arrival influences the performance of network MIMO requires further study. Specifically, as the JT is constrained by the per-BS power budget, the performance of the network MIMO is limited if the power budgets are severely asymmetric among BSs. For example, if a solar-powered BS in a windless sunny day cooperates with a wind-powered BS, the latter will become the performance bottleneck of cooperation, while the harvested energy of the former is not efficiently utilized. To deal with this problem, people have introduced the concept of energy cooperation \cite{gurakan2013energy, chia2014energy}, where BSs can exchange energy via either wired or wireless link with some loss of energy transfer. In this case, the JT problem with energy harvesting becomes a power allocation problem with weighted sum power constraint as shown in \cite{jie2015comp}. However, the feasibility and efficiency of cooperation in energy domain strongly depends on the existence and the efficiency of energy transfer link.

In this paper, we consider how to improve the utilization of harvested energy with cooperation between the wireless radio links. Intuitively, if the energy cannot be transferred between BSs, {the BS with higher energy arrival rate should use more energy in data transmission to avoid energy waste. While to use the energy more effective, BS cooperation strategy should be carefully designed under the asymmetric energy constraints.} Based on this, we propose a \emph{fractional JT} strategy, where the network MIMO is only applied in a fraction of a transmission frame. Specifically, we consider two BSs cooperatively serving two users, and divide each transmission frame into two subframes. In the first subframe, one of the BSs serves one user while the other stores energy. In the second subframe, the two BSs perform JT to cooperatively serve the two users. With the stored energy, the power gap between two BSs in the second subframe is filled, and hence, JT can achieve higher sum-rate. Such a strategy avoids the potential energy waste in the BS with higher energy arrival rate, and hence can improve the energy utilization. The objective is to maximize the average sum-rate for given energy arrival rates, and the optimization parameters include the fraction of time for JT and the power allocation policy in each frame. Our preliminary work \cite{gong2014downlink} has studied the greedy policy that tries to use all the available energy in each frame. In this paper, we further consider the optimal policy as well as the low-complexity policy. The contributions of this paper are as follows.

\begin{itemize}
 \item We propose the fractional JT strategy, and formulate the long-term average sum-rate maximization problem using Markov decision process (MDP) \cite{bertsekas2005dynamic}. The problem is divided into two sub-problems, i.e., energy management among frames, and power allocation problem for fractional JT in each frame.
 \item We prove that to solve the average sum-rate maximization problem, in each frame, we only need to solve the power allocation problem with equality power constraints, which has closed-form expressions. Then the JT time fraction decision problem is proved to be a convex optimization problem, and a bi-section search algorithm is proposed to find the optimal JT time fraction.
 \item We adopt the \emph{approximate dynamic programming} (DP) \cite{bertsekas2005dynamic} algorithm to reduce the computational complexity of determining the energy allocation among frames. The algorithm runs iteratively with two steps: \emph{policy evaluation} and \emph{policy improvement}. In the policy evaluation, the relative utility function in the Bellman's equation is approximated as a weighted summation of a set of features associated with system states. The weights are estimated by simulation. In the policy improvement, random initial policies are periodically selected to rerun the iteration to broaden the search range. Numerical simulations show the remarkable performance gain compared with the conventional network MIMO.
\end{itemize}

The rest of the paper is organized as follows. Section \ref{sec:model} describes the system model and Section \ref{sec:problem} describes the MDP problem formulation. In Section \ref{sec:perstage}, the per-frame optimization problem is analyzed. Then the approximate DP algorithm is proposed in Section \ref{sec:ADP}. Simulation study is presented in Section \ref{sec:sim}. Finally, Section \ref{sec:concl} concludes the paper.

\emph{Notations:} Bold upper case and lower case letters denote matrices and vectors, respectively. $|\cdot|$ denotes the absolute value of a scalar, and $[x]^+ = \max\{x, 0\}$. $(\cdot)^T$ and $(\cdot)^H$ denote the transpose and transpose conjugate of a matrix, respectively. ${\cal R}^{+}$ is the non-negative real number field. $\mathbb{E}$ represents the expectation operation.

\section{System Model} \label{sec:model}
\begin{figure}
\centering
\includegraphics[width=4.2in]{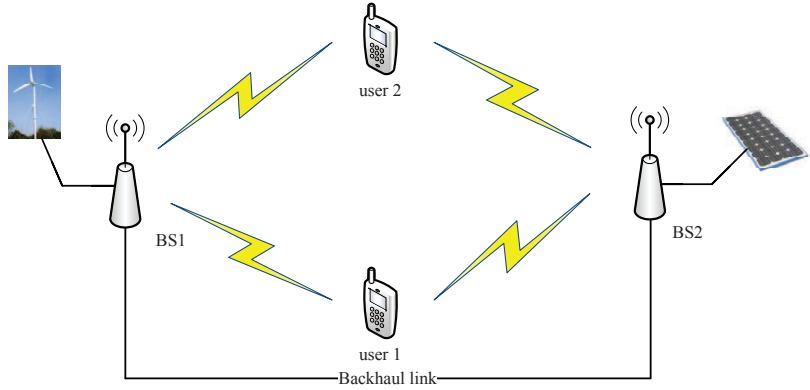}
\caption{System model for joint transmission with 2 BSs and 2 users.} \label{fig:system}
\end{figure}

We consider a wireless communication network consisting of two BSs powered by renewable energy (e.g., solar energy, wind energy, etc.) and two users as shown in Fig.~\ref{fig:system}. Assume the BSs are able to store the harvested energy in their battery for future usage. All the BSs and the users are equipped with a single antenna. The BSs are interconnected via an error-free backhaul link sharing all the data and the channel state information, so that they can perform JT to eliminate the interference. { However, the energy cannot be transferred between the BSs as we consider the off-grid scenario.} We consider the typical scenario for applying network MIMO, in which the two users are located at the cell boundary. In this case, the average channel gains are comparable, and hence cooperative transmission can achieve significant performance gain. The wireless channel is assumed block fading, i.e., the channel state is constant during each fading block, but changes from block to block. We define the transmission frame as a channel fading block with frame length $T_f$. The perfect channel state information is assumed known to the BSs at the beginning of each frame. { If the backhaul capacity is limited, the two BSs can exchange quantized data and channel state information, and cooperate in the same way using the imperfect information.}

In the $t$-th frame, if the JT technique is utilized, the received signals $\mathbf{y}_t = [y_{t,1}, y_{t,2}]^T$ at the users are
\begin{equation}
\mathbf{y}_t = \mathbf{H}_t\mathbf{W}_t\mathbf{x}_t + \mathbf{n}_t,
\end{equation}
where $\mathbf{H}_t$ is the channel matrix with components $H_{t,ik} = l_{ik}\tilde{H}_{t,ik}, 1\le i,k \le 2$ indicating the channel coefficient from BS $k$ to user $i$ with large-scale fading factor $l_{ik}$ and i.i.d. small-scale fading factor $\tilde{H}_{t,ik}$, $\mathbf{W}_t$ is the corresponding precoding matrix with components $w_{t,ki}$, $\mathbf{x}_t = [x_{i,1}, x_{t,2}]^T$ is the intended signals for the users with $\mathbb{E}(\mathbf{x}_t\mathbf{x}_t^H) = \mathrm{diag}(p_{t,1}, p_{t,2})$, where $p_{t,i}, i = 1, 2$ is the power allocated to user $i$, and $\mathbf{n}_t$ is the additive white Gaussian noise with zero mean and variance $\mathbb{E}(\mathbf{n}_t\mathbf{n}_t^H) = \sigma_n^2\mathbf{I}$, where $\mathbf{I}$ is a $2\times 2$ unit matrix.

{In this paper, the widely used ZF precoding scheme \cite{boccardi2006zero} is adopted to completely eliminate the interference by channel inverse. Thus, the decoding process at the users can be simplified. And its performance can be guaranteed, especially when the interference dominates the noise. In addition, ZF precoding is a representative precoding scheme. Hence, the following analysis can be easily extended to other schemes.} For ZF precoding scheme, we have
\begin{equation}
\mathbf{W}_t = \mathbf{H}_t^{-1}.
\end{equation}
Hence, the data rate is
\begin{equation}
R_{t,i} = \log_2(1+\frac{p_{t,i}}{\sigma^2_n})\label{eq:Rcomp}
\end{equation}
with per-BS power constraint
\begin{equation}
\sum_{i=1}^2 |w_{t,ki}|^2 p_{t,i} \le P_{t,k}, \quad k = 1, 2. \label{eq:powerctr}
\end{equation}
where $P_{t,k}$ is the maximum available transmit power of BS $k$ in frame $t$. {Notice that if the BSs and the users are equipped with multiple antennas, ZF precoding scheme should be replaced by the multi-cell block diagonalization (BD) \cite{zhang2009networked} scheme which also nulls the inter-BS interference. As the multi-cell BD scheme is a generalization of ZF precoding scheme from single antenna case to multi-antenna case, it has similar mathematical properties with the latter. Hence, the following results can be extened to multi-antenna case.}

As the BSs are powered by the renewable energy, $P_{t,k}$ is determined by the amount of harvested energy as well as the available energy in the battery. It is pointed out in \cite{chuan2013threshold, chuan2014optimal} that in real systems, the energy harvesting rate changes in a much slower speed than the channel fading. Specifically, a fading block in current wireless communication systems is usually measured in the time scale of milliseconds, while the renewable energy such as solar power may keep constant for seconds or even minutes. Hence, the energy arrival rate (energy harvesting power) is assumed constant over a sufficient number of transmission frames, denoted by $E_k, k = 1, 2$. In this case, the key factor of the energy harvesting is the energy arrival causality constraint, i.e., the energy that has not arrived yet cannot be used in advance. In this paper, we mainly study the influence of the energy causality on the network MIMO.

{Notice that in practice, the optimization over multiple energy coherence blocks is required as the energy arrival rate varies over time. If the future energy arrival information is unknown (i.e., purely random and unpredictable), we can monitor the energy harvesting rate and once it changes, we recalculate the optimal policy under the new energy constraint, and then apply the new policy. The policy optimization problem is considered in this paper. While if the energy arrival rate is predictable, the optimization should jointly consider multiple blocks in the prediction window, which is beyond the scope the this paper.}

\subsection{Fractional Joint Transmission Strategy}
Notice that the energy arrival rates of different BSs may be different due to either utilizing various energy harvesting equipments (e.g., one with solar panel, the other with wind turbine) or encountering different environment conditions (e.g., partly cloudy). In this case, the conventional network MIMO may not be sum-rate optimal as the harvested energy is not efficiently utilized. Specifically, if the channel conditions of the two users are similar, applying network MIMO with on average the same energy usage can achieve the optimal cooperation efficiency. As a result, in the asymmetric energy arrival case, the energy of the BS with higher energy arrival rate may be not efficiently used. Hence, the performance of network MIMO may be greatly degraded. {Notice that the above fact does not only hold for ZF precoding, but also holds for other approaches (such as the approach based on dirty paper coding \cite{caire2003achievable, karakayali2007network}) as it is caused by the asymmetric per-BS power constraints, rather than the precoding scheme itself.}

\begin{figure}
\centering
\includegraphics[width=3.2in]{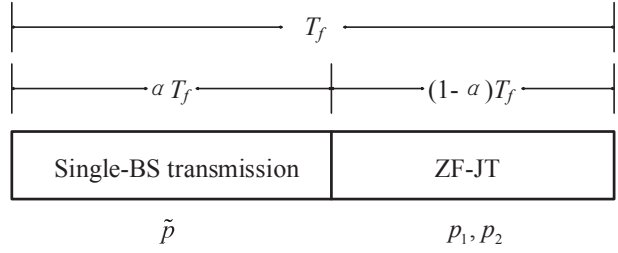}
\caption{Frame structure of fractional JT. The frame length is $T_f$.} \label{fig:frame}
\end{figure}

To improve the utilization of the harvested energy, we propose a fractional JT strategy to adapt to the asymmetric energy arrival rates. Thanks to the energy storage ability, the BS can turn to sleep mode to store energy for a while, and then cooperatively transmits data with the other BS. In this way, it can provide higher transmit power when applying network MIMO. The strategy is detailed as follows. We divide the whole transmission frame into two subframes as shown in Fig.~\ref{fig:frame}. In the first subframe, named as \emph{single-BS transmission phase}, one of the BSs $k_t \in \{1,2\}$ is selected to serve a user, while the other one, denoted by $\bar{k}_t \neq k_t$, turns to sleep mode to store energy. In the second subframe, named as \emph{ZF-JT phase}, the two BSs jointly transmit to the two users with ZF precoding scheme as explained earlier in this section. Denote by $\alpha_tT_f$ the length of the single-BS transmission phase, where $0 \le \alpha_t \le 1$, and hence, the length of the ZF-JT phase is $(1-\alpha_t)T_f$. To get the optimal fractional JT transmission strategy, we need to choose $k_t$ and $\alpha_t$ carefully.

In the single-BS transmission phase, to be consistent with the objective of maximizing sum-rate, the active BS serves one of the users with higher instantaneous data rate. Specifically, the user $\tilde{i}$ is scheduled when satisfying $\tilde{i} = \arg \max_{1\le i \le 2} \log_2(1+ \frac{\bar{P}|H_{t,ik_t}|^2}{\sigma^2_n}),$ i.e., the user with the maximum expected data rate with transmit power $\bar{P}= E_{k_t}$. {In practice, the proposed fractional JT transmission strategy can be supported by the CoMP \cite{3GPP2011TR}, in which all the data is shared by the two BSs in both subframes. Notice that as only one BS is active in the first subframe, the data transferred to the inactive BS via the backhaul is useless, and such a backhaul data sharing protocol is inefficient.}

{However, when the backhaul capacity is limited, the proposed fractional JT strategy can make use of the backhaul capacity in the first subframe to enhance the performance. Since the shared data is required only in the second subframe, the two BSs  in the first subframe can proactively exchange the data to be jointly transmitted later. Thus, the quantization noise of the shared data can be reduced and the cooperation gain can be enhanced.}

\subsection{Sum-rate Maximization Problem}
Our objective is to optimize the sum-rate under the proposed fractional JT strategy. The power constraints in each frame are detailed as follows. The available energy in the battery of the active BS $k_t$ at the beginning of each frame $t$ is denoted by $B_{t,k_t}$. Then the power in the first subframe satisfies
\begin{equation}
\tilde{p}_{t} \le \frac{B_{t,k_t}}{\alpha_t T_f} + E_{k_t}. \label{eq:power1}
\end{equation}
At the beginning of the second subframe, the amounts of available battery energy in the two BSs become $B_{t,k_t}+\alpha_t T_fE_{k_t} -\alpha_t T_f\tilde{p}_{t}$ and $B_{i,\bar{k}_t}+\alpha_t T_fE_{\bar{k}_t}$, respectively. As a result, the power constraints (\ref{eq:powerctr}) for ZF-JT become
\begin{align}
& \sum_{i=1}^2 |w_{t,k_ti}|^2 p_{t,i} \le \frac{B_{t,k_t}+\alpha_t T_f(E_{k_t} -\tilde{p}_{t})} {(1-\alpha_t)T_f} + E_{k_t}, \label{eq:power2}\\
& \sum_{i=1}^2 |w_{t,\bar{k}_ti}|^2 p_{t,i} \le \frac{B_{t,\bar{k}_t}+\alpha_t T_fE_{\bar{k}_t}}{(1-\alpha_t)T_f} + E_{\bar{k}_t}. \label{eq:power3}
\end{align}
The battery energy states are updated according to
\begin{align}
& B_{t\!+\!1\!,k_{t+1}} \!=\! B_{t,k_t} \!+\! T_f(E_{k_t} \!-\! \alpha_t\tilde{p}_{t} \!-\! (1\!-\!\alpha_t) \sum_{i=1}^2 |w_{t,k_ti}|^2 p_{t,i}), \label{eq:battery1}\\
& B_{t\!+\!1\!,\bar{k}_{t+1}} \!=\! B_{t,\bar{k}_t} \!+\! T_f(E_{\bar{k}_t} \!-\! (1\!-\!\alpha_t)\sum_{i=1}^2 |w_{t,\bar{k}_ti}|^2 p_{t,i}), \label{eq:battery2}
\end{align}
with initial state $B_{1,1} = B_{1,2} = 0$. In (\ref{eq:power1}), (\ref{eq:power2}), and (\ref{eq:power3}), we have $0<\alpha_t<1$ as the denominator cannot be zero. In fact, by multiplying $\alpha_t$ on both sides of (\ref{eq:power1}) and $1-\alpha_t$ on both sides of (\ref{eq:power2}) and (\ref{eq:power3}), the special case that $\alpha_t = 0 \textrm{~or~} 1$ can be included in a unified formulation. Denote by $\mathbf{{k}} = \{k_1, k_2, \cdots, k_N\}$, $\mathbf{{\alpha}} = \{\alpha_1, \alpha_2, \cdots, \alpha_N\}$, $\mathbf{{\tilde{p}}} = \{\tilde{p}_1, \tilde{p}_2, \cdots, \tilde{p}_N\}$, $\mathbf{{p}} = \{\mathbf{\emph{p}}_1, \mathbf{\emph{p}}_2, \cdots, \mathbf{\emph{p}}_N\}$, where $\mathbf{\emph{p}}_t = (p_{t,1}, p_{t,2})^T$, and $N$ is the number of transmission frames. Our optimization problem can be formulated as
\begin{align}
\max_{\mathbf{k}, \mathbf{\alpha}, \mathbf{\tilde{p}}, \mathbf{p}} \;& \lim_{N\rightarrow \infty}\mathbb{E}_{\mathbf{H}}\!\left[\!\frac{1}{N}\sum_{t=1}^N \left(\alpha_t\tilde{R}_{t,\tilde{i}} \!+\! (1\!-\!\alpha_t)\sum_{i=1}^{2}R_{t,i}\right)\!\right] \label{eq:problem}\\
\mathrm{s.t.~}&\alpha_t\tilde{p}_t \le \frac{B_{t,k_t}}{T_f} + \alpha_tE_{k_t}, \label{eq:power1o}\\
 & (1\!-\!\alpha_t) \sum_{i=1}^2 |w_{t,k_ti}|^2 p_{t,i} \!+\! \alpha_t\tilde{p}_t \le \frac{B_{t,k_t}}{T_f}\!+\!E_{k_t}, \label{eq:power2o}\\
& (1-\alpha_t)\sum_{i=1}^2 |w_{t,\bar{k}_ti}|^2 p_{t,i} \le \frac{B_{t,\bar{k}_t}}{T_f} + E_{\bar{k}_t}, \label{eq:power3o}\\
 {} & \tilde{p}_t, p_{t, 1}, p_{t, 2} \in {\cal R}^{+}, \qquad \forall t = 1, 2, \cdots, N. \label{eq:power4o}\\
 {} & 0 \le \alpha_t \le 1, \label{eq:power5o}
\end{align}
where $\tilde{R}_{t,\tilde{i}} = \log_2(1+\tilde{p}_t|H_{t,\tilde{i}k_t}|^2/\sigma^2_n)$, and $R_{t,i}$ is expressed as (\ref{eq:Rcomp}). The optimization parameters include the transmit power $\tilde{p}_t, p_{t,k}, k = 1, 2$, the frame division parameter $\alpha_t$, and the selection of BSs $k_t$ for single-BS transmission phase. Notice that if $\alpha_t=0$, the problem reduces to the conventional power allocation problem for network MIMO; if $\alpha_t=1$, the problem becomes user selection and rate maximization problem for single-BS transmission. To find the optimal solution, we need to calculate the integration of the channel distribution over all the frames and exhaustively search all the possible power allocation and frame division policies, which is computationally overwhelming. In the work, we aim to design a low-complex algorithm to achieve close-to-optimal performance.

\section{MDP Modeling and Optimization} \label{sec:problem}
In this section, we reformulate the stochastic optimization problem (\ref{eq:problem}) based on the MDP framework \cite{bertsekas2005dynamic}. Specifically, in each channel fading block, we need to decide which BS should turn to sleep to store energy in the first subframe, how long it should sleep, and how much power should be allocated. The decision in each frame will influence the decisions in the future, as it changes the remained energy in the battery. MDP is an effective mathematical framework to model such a time-correlated decision making problem. The formulation is detailed as follows.

\subsection{MDP Problem Reformulation}
A standard MDP problem contains the following elements: state, action, per-stage utility function and state transition. In our problem, the stage refers to the frame. In each stage, the system state includes the battery states of two BSs at the beginning of the frame and the channel states, i.e., $s_t = (B_{t,1}, B_{t, 2}, \mathbf{H}_t)$. Denote the state space by $\mathcal{S}$. We model the action as the power budget of each frame, i.e., $a_t(s_t) = (A_{t,1}, A_{t,2})$ which satisfies $0 \le A_{t,1} \le \frac{B_{t,1}}{T_f} + E_1$ and $0 \le A_{t,2} \le \frac{B_{t,2}}{T_f} + E_2.$
We denote the state-dependent action space by $\mathcal{A}(s_t) = \{(A_{t,1}, A_{t,2})|0 \le A_{t,1} \le \frac{B_{t,1}}{T_f} + E_1, 0 \le A_{t,2} \le \frac{B_{t,2}}{T_f} + E_2 \}$.
%\begin{align}
%\mathcal{A}(s_t) = \{(A_{t,1}, A_{t,2})|&0 \le A_{t,1} \le \frac{B_{t,1}}{T_f} + E_1,\nonumber\\
%{ }& 0 \le A_{t,2} \le \frac{B_{t,2}}{T_f} + E_2 \}.
%\end{align}
The per-stage sum-rate function can be expressed as
\begin{equation}
g(s_t, a_t) = \max_{k_t\!, \alpha_t\!, \tilde{p}_t\!, \mathbf{\emph{p}}_t} \alpha_t\!\log_2\!\Big(\!1\!+\!\frac{\tilde{p}_t|H_{t,\tilde{i}k_t}|^2}{\sigma^2_n}\!\Big) \!+ \! (\!1\!-\!\alpha_t\!)\sum_{i=1}^{2}\log_2\Big(\!1\!+\!\frac{p_{t,i}}{\sigma^2_n}\!\Big), \label{eq:frameprobg}
\end{equation}
%\begin{align}
%&g(s_t, a_t) = \nonumber\\
%&\max_{k_t\!, \alpha_t\!, \tilde{p}_t\!, \mathbf{\emph{p}}_t} \alpha_t\!\log_2\!\Big(\!1\!+\!\frac{\tilde{p}_t|H_{t,\tilde{i}k_t}|^2}{\sigma^2_n}\!\Big) \!+ \! (\!1\!-\!\alpha_t\!)\sum_{i=1}^{2}\log_2\Big(\!1\!+\!\frac{p_{t,i}}{\sigma^2_n}\!\Big), \label{eq:frameprobg}
%\end{align}
where the maximization is taken under the constraints (\ref{eq:power1o}), (\ref{eq:power4o}), (\ref{eq:power5o}) and
\begin{align}
 & (1-\alpha_t) \sum_{i=1}^2 |w_{t,k_ti}|^2 p_{t,i} + \alpha_t\tilde{p}_t \le A_{t,k_t}, \label{eq:power2g}\\
& (1-\alpha_t)\sum_{i=1}^2 |w_{t,\bar{k}_ti}|^2 p_{t,i} \le A_{t, \bar{k}_t}, \label{eq:power3g}
\end{align}

The state transition of the battery energy is deterministic according to (\ref{eq:battery1}) and (\ref{eq:battery2}). The channel state of the next stage is obtained according to the channel transition $\mathrm{Pr}(\mathbf{H}_{t+1}|\mathbf{H}_t)$, which is independent with the battery energy state.

Consequently, the original problem (\ref{eq:problem}) can be reformulated as
\begin{align}
\max_{\bm{a}} \;& \lim_{N\rightarrow \infty}\mathbb{E}_{\mathbf{H}}\!\left[\!\frac{1}{N}\sum_{t=1}^N g(s_t, a_t(s_t))\!\right]. \label{eq:problemMDP}
\end{align}
The optimization is taken over all the possible policies $\bm{a} = \{a_1, a_2, \ldots\}$. It is obvious that for any two states, there is a stationary policy $\bm{a}$ so that one state can be accessed from the other with finite steps \cite[Sec~4.2]{bertsekas2005dynamic}. Consequently, the optimal value is independent of the initial state and there exists an optimal stationary policy $\bm{a}^* = \{a^*(s)|s \in \mathcal{S}\}$ .

\subsection{Value Iteration Algorithm}
According to \cite[Prop.~4.2.1]{bertsekas2005dynamic}, there exists a scalar $\Lambda^*$ together with some vector $\bm{h}^* = \{h^*(s)|s \in \mathcal{S}\}$ satisfies the Bellman's equation
\begin{equation}
\Lambda^* + h^*(s) = \max_{a\in \mathcal{A}(s)}\left[ g(s, a) + \sum_{s' \in \mathcal{S}}p_{s\rightarrow s' | a}h^*(s')\right], \label{eq:bellman}
\end{equation}
where $\Lambda^*$ is the optimal average utility, and $h^*(s)$ is viewed as \emph{relative or differential utility}\footnote{In the textbook \cite{bertsekas2005dynamic}, $h^*(s)$ is defined as \emph{relative cost} instead since the objective there is to minimize the average cost}. It represents the maximum difference between the expected utility to reach a given state $s_0$ from state $s$ for the first time and the utility that would be gained if the utility per stage was the average $\Lambda^*$. Furthermore, if $a^*(s)$ attains the maximum value of (\ref{eq:bellman}) for each $s$, the stationary policy $\bm{a}^*$ is optimal. Based on the Bellman's equation, instead of the long term average sum-rate maximization, we only need to deal with (\ref{eq:bellman}) which only relates with per-stage sum-rate $g(a, s)$ and state transition $p_{s\rightarrow s' | a}$. The value iteration algorithm \cite[Sec.~4.4]{bertsekas2005dynamic} can effectively solve the problem.

Specifically, we firstly initialize $h^{(0)}(s) = 0, \forall s \in \mathcal{S}$, and set a parameter $0 < \tau < 1$, which is used to guarantee the convergence of value iteration while obtaining the same optimal solution \cite[Prop. 4.3.4]{bertsekas2005dynamic}. Then we choose a state to calculate the relative utility. We choose a fixed state $s_0 = (0, 0, \mathbf{H}_0)$, and denote the output of the $n$-th iteration as $\bm{h}^{(n)}= \{h^{(n)}(s)|s \in \mathcal{S}\}$. For the $(n\!+\!1)$-th iteration, we first calculate
\begin{equation}
\Lambda^{(n+1)}(s_0) = \max_{a\in \mathcal{A}(s_0)}\left[ g(s_0, a) + \tau\sum_{ \mathbf{H}'} \mathrm{Pr} (\mathbf{H}'|\mathbf{H_0})h^{(n)}(s')\right], \label{eq:lambdanplus1}
\end{equation}
where $s' = (B_1', B_2', \mathbf{H}')$, and $B_1', B_2'$ are calculated according to (\ref{eq:battery1}) and (\ref{eq:battery2}), respectively. Then we calculate the relative utilities as
\begin{equation}
h^{(n+1)}(s) = (1-\tau)h^{(n)}(s) + \max_{a\in \mathcal{A}(s)}\left[ g(s, a) + \tau\sum_{ \mathbf{H}'} \mathrm{Pr} (\mathbf{H}'|\mathbf{H})h^{(n)}(s')\right] - \Lambda^{(n+1)}(s_0). \label{eq:hnplus1}
\end{equation}
%\begin{align}
%&h^{(n+1)}(s) = (1-\tau)h^{(n)}(s) + \nonumber\\
%&\max_{a\in \mathcal{A}(s)}\left[ g(s, a) + \tau\sum_{ \mathbf{H}'} \mathrm{Pr} (\mathbf{H}'|\mathbf{H})h^{(n)}(s')\right] - \Lambda^{(n+1)}(s_0). \label{eq:hnplus1}
%\end{align}
Recall that the parameter $\tau$ is used to guarantee the convergence of the relative value iteration. It can be viewed as replacing the relative utility $h(s)$ by $\tau h(s)$, which is proved not to change the optimal value. As the optimal average utility is irrelative with the initial state, $\Lambda^{(n+1)}(s_0)$ converges to $\Lambda^*$.

Notice that the states and the actions are all in the continuous space. By discretizing the state space and the action space, the MDP framework can be applied to solve the problem. However, to make the solution accurate, the granularity of the discretization needs to be sufficiently small, which results in a tremendous number of states, especially for the $2\times2$ MIMO channels (4 elements, each with two scalars: real part and imaginary part). As a consequence, we need to not only calculate the per-stage sum-rate function $g(s, a)$ that includes maximization operation for all states, but also iteratively update all the relative utilities $h(s)$. In this sense, solving the MDP problem encounters unaffordable high computational complexity, which is termed as the \emph{curse of dimensionality} \cite{bertsekas2005dynamic}. To reduce the computational complexity, on the one hand, the maximization problem in the per-stage sum-rate function should be solved efficiently. On the other hand, the complexity of the iteration algorithm should be reduced via some approximation. In the next two sections, we will discuss these two aspects in detail.

\section{Per-Frame Sum-Rate Maximization} \label{sec:perstage}
In this section, we firstly consider the per-stage sum-rate function $g(s_t, a_t)$, i.e., the sum-rate maximization problem in each frame for the current state $s_t = (B_{t,1}, B_{t, 2}, \mathbf{H}_t)$ and the given action $a_t = (A_{t,1}, A_{t,2})$. We ignore the time index $t$ for simplicity. The per-frame optimization problem can be formulated as
\begin{align}
\max_{k\!, \alpha\!, \tilde{p}\!, p_1\!, p_2\!}  \quad& \alpha\!\log_2\!\Big(\!1\!+\!\frac{\tilde{p}|H_{\tilde{i}k}|^2}{\sigma^2_n}\!\Big) \!+ \! (\!1\!-\!\alpha\!)\sum_{i=1}^{2}\log_2\Big(\!1\!+\!\frac{p_{i}}{\sigma^2_n}\!\Big) \label{eq:frameprob}\\
\mathrm{s.t.~}\quad&\alpha\tilde{p} \le \frac{B_{k}}{T_f} + \alpha E_{k}, \label{eq:power1p}\\
 & (1-\alpha) \sum_{i=1}^2 |w_{ki}|^2 p_{i} + \alpha\tilde{p} \le A_{k}, \label{eq:power2p}\\
& (1-\alpha)\sum_{i=1}^2 |w_{\bar{k}i}|^2 p_{i} \le A_{\bar{k}}, \label{eq:power3p}\\
 {} & \tilde{p}, p_{1}, p_{2} \in {\cal R}^{+}. \label{eq:power4p}\\
 {} & 0 \le \alpha \le 1. \label{eq:power5p}
\end{align}
As $k \in \{1, 2\}$, the optimization over $k$ can be done by solving the problem for all $k$, and selecting the one with larger sum-rate. Thus, we only need to consider the problem for a given $k$. %Without loss of generality, we assume $k = 1, \bar{k} = 2$, i.e., BS2 turns to sleep mode in the single-BS transmission phase.
Then the optimization problem can be rewritten as
\begin{align}
\max_{\alpha\!, \tilde{p}\!, p_1\!, p_2\!}  \quad& \alpha\!\log_2\!\Big(\!1\!+\!\frac{\tilde{p}|H_{\tilde{i}k}|^2}{\sigma^2_n}\!\Big) \!+ \! (\!1\!-\!\alpha\!)\sum_{i=1}^{2}\log_2\Big(\!1\!+\!\frac{p_{i}}{\sigma^2_n}\!\Big) \label{eq:frameprobk}%\\
%\mathrm{s.t.~}\quad&(\ref{eq:power1p}), (\ref{eq:power2p}), (\ref{eq:power3p}), \textrm{~and~} (\ref{eq:power4p}) \nonumber
\end{align}
The problem (\ref{eq:frameprobk}) with constraints (\ref{eq:power1p})-(\ref{eq:power5p}) is not convex in general. However, as shown later, given $\alpha$, the power allocation problem is a convex optimization, and the optimization over $\alpha$ given the optimal power allocation is also convex. According to these properties, we study the optimization of power allocation and subframe division separately.

\subsection{Power Allocation Optimization}
If we fix the variables $k$ and $\alpha$ in (\ref{eq:frameprobk}), we obtain a power allocation optimization problem, which has the following property.
\begin{theorem}\label{prop:convex}
For given $k$ and $\alpha$, the problem
\begin{align}
\max_{\tilde{p}\!, p_1\!, p_2\!}  \quad& \alpha\!\log_2\!\Big(\!1\!+\!\frac{\tilde{p}|H_{\tilde{i}k}|^2}{\sigma^2_n}\!\Big) \!+ \! (\!1\!-\!\alpha\!)\sum_{i=1}^{2}\log_2\Big(\!1\!+\!\frac{p_{i}}{\sigma^2_n}\!\Big) \label{prob:poweralloc}
\end{align}
with constraints (\ref{eq:power1p}) - (\ref{eq:power4p}) is a convex optimization problem.
\end{theorem}
\begin{IEEEproof}
Once $\alpha$ is fixed, the objective function is the maximization of a summation of concave functions, and all the constraints are linear. As a result, the problem is convex.
\end{IEEEproof}

Theorem \ref{prop:convex} tells us that for a given $k$ and $\alpha$, the optimal solution can be found by solving a convex optimization problem for power allocation. According to the convex optimization theory \cite{boyd2004convex}, we have the following observation.

%A quick observation of the above problem is stated as follows:
\begin{proposition} \label{prop:greedy}
For a given $k$, when the optimal solution for the problem (\ref{eq:frameprobk}) with constraints (\ref{eq:power1p})-(\ref{eq:power5p}) is achieved, either (\ref{eq:power2p}) or (\ref{eq:power3p}) is satisfied with equality.
\end{proposition}
\begin{IEEEproof}
See Appendix \ref{proof:greedy}.
\end{IEEEproof}

However, Proposition \ref{prop:greedy} cannot guarantee the equality holds in both (\ref{eq:power2p}) and (\ref{eq:power3p}). If both are satisfied with equality, the problem can be simplified and the solution can be given in closed-form. As a matter of fact, an equivalent problem can be formulated which only needs to solve the power allocation problem with equality held in (\ref{eq:power2p}) and (\ref{eq:power3p}). To get the result, we firstly provide a useful lemma as follows.

\begin{lemma} \label{lemma:hincre}
The relative utility $h^*(s) = h^*(B_1, B_2, \mathbf{H})$ is nondecreasing w.r.t. $B_1$(or $B_2$) for given $B_2$(or $B_1$) and $\mathbf{H}$.
\end{lemma}
\begin{IEEEproof}
See Appendix \ref{proof:hincre}.
\end{IEEEproof}

Intuitively, more energy in the battery can support higher data rate. Hence, the utility increases with the increase of the battery energy. Based on Lemma \ref{lemma:hincre}, we have the following conclusion.

\begin{theorem} \label{prop:gbar}
Define $\bar{g}(s, a) = g(s, a)$ where the optimization is under the constraints (\ref{eq:power1p}), (\ref{eq:power4p}), (\ref{eq:power5p}) and the equality constraints
\begin{align}
 & (1-\alpha) \sum_{i=1}^2 |w_{ki}|^2 p_{i} + \alpha\tilde{p} = A_{k}, \label{eq:power2gb}\\
& (1-\alpha)\sum_{i=1}^2 |w_{\bar{k}i}|^2 p_{i} = A_{\bar{k}}, \label{eq:power3gb}
\end{align}
we have
\begin{align}
\Lambda^* = \max \; \lim_{N\rightarrow \infty}\mathbb{E}_{\mathbf{H}}\!\left[\!\frac{1}{N}\sum_{t=1}^N g(s_t, a_t(s_t))\!\right] = \max \; \lim_{N\rightarrow \infty}\mathbb{E}_{\mathbf{H}}\!\left[\!\frac{1}{N}\sum_{t=1}^N \bar{g}(s_t, a_t(s_t))\!\right] \nonumber
\end{align}
%\begin{align}
%\Lambda^* &= \max \; \lim_{N\rightarrow \infty}\mathbb{E}_{\mathbf{H}}\!\left[\!\frac{1}{N}\sum_{t=1}^N g(s_t, a_t(s_t))\!\right] \nonumber\\
%{ }&= \max \; \lim_{N\rightarrow \infty}\mathbb{E}_{\mathbf{H}}\!\left[\!\frac{1}{N}\sum_{t=1}^N \bar{g}(s_t, a_t(s_t))\!\right] \nonumber
%\end{align}
\end{theorem}
\begin{IEEEproof}
See Appendix \ref{proof:gbar}.
\end{IEEEproof}

Based on Theorem \ref{prop:gbar}, we only need to solve the maximization problem under the equality constraints (\ref{eq:power2gb}) and (\ref{eq:power3gb}). The optimal power allocation solution as follows.

\begin{proposition} \label{prop:ptilde}
For a given $k$ and $0 < \alpha < 1$, we denote
\begin{align}
\tilde{p}_{\mathrm{min}} &= \max\Big\{0, \frac{C_{2}}{\alpha|w_{\bar{k}1}|^2}\Big\}, \label{ptilde:min}\\
\tilde{p}_{\mathrm{max}} &= \min\Big\{\frac{B_k}{\alpha T_f} + E_k, \frac{C_{1}}{\alpha|w_{\bar{k}2}|^2}\Big\}, \label{ptilde:max}
\end{align}
define the set ${\cal P}_{k,\alpha} = \left\{\tilde{p} \Big| \tilde{p}_{\mathrm{min}} \le \tilde{p} \le \right. \left.\tilde{p}_{\mathrm{max}} \right\}$, and denote $\tilde{p}_0$ as the nonnegative root of
\begin{align}
\frac{|H_{\tilde{i}k}|^2}{\sigma^2_n + \tilde{p}|H_{\tilde{i}k}|^2} - \frac{(1-\alpha)  |w_{\bar{k}2}|^2}{\sigma^2_nC_0 + C_1 - \alpha |w_{\bar{k}2}|^2\tilde{p}} + \frac{(1-\alpha)  |w_{\bar{k}1}|^2}{\sigma^2_nC_0 + C_2 + \alpha |w_{\bar{k}1}|^2\tilde{p}} = 0,
\label{eq:quadratic}
\end{align}
%\begin{align}
%\frac{|H_{\tilde{i}k}|^2}{\sigma^2_n + \tilde{p}|H_{\tilde{i}k}|^2} - \frac{(1-\alpha)  |w_{\bar{k}2}|^2}{\sigma^2_nC_0 + C_1 - \alpha |w_{\bar{k}2}|^2\tilde{p}} + &{}\nonumber\\
%\frac{(1-\alpha)  |w_{\bar{k}1}|^2}{\sigma^2_nC_0 + C_2 + \alpha |w_{\bar{k}1}|^2\tilde{p}} &= 0 \label{eq:quadratic}
%\end{align}
where $C_0 = (1-\alpha)(|w_{k1}|^2|w_{\bar{k}2}|^2 - |w_{k2}|^2|w_{\bar{k}1}|^2),
C_{1} = A_k|w_{\bar{k}2}|^2 - A_{\bar{k}}|w_{k2}|^2,
C_{2} = A_{k}|w_{\bar{k}1}|^2  - A_{\bar{k}}|w_{k1}|^2.$ Then the solution for the problem (\ref{prob:poweralloc}) with constraints (\ref{eq:power1p}), (\ref{eq:power4p}), (\ref{eq:power2gb}) and (\ref{eq:power3gb}) is

\begin{itemize}
\item If ${\cal P}_{k,\alpha} = \emptyset$, the problem is infeasible.

\item Otherwise, we have

(1) if $\tilde{p}_0 \in {\cal P}_{k,\alpha}$, $\tilde{p}^* = \tilde{p}_0$ is the optimal power for the single-BS transmission phase;

(2) if $\tilde{p}_0 > \tilde{p}_{\mathrm{max}}$, $\tilde{p}^* = \tilde{p}_{\mathrm{max}}$ is optimal;

(3) if $\tilde{p}_0 < \tilde{p}_{\mathrm{min}}$, $\tilde{p}^* = \tilde{p}_{\mathrm{min}}$ is optimal;

and the optimal $p_i^*, i = 1, 2$ can be obtained via
\begin{align}
p_1^* = & \frac{C_{1} - \alpha|w_{\bar{k}2}|^2\tilde{p}^*}{C_0}, \label{eq:p1star}\\
p_2^* = & \frac{\alpha|w_{\bar{k}1}|^2\tilde{p}^* - C_{2}}{C_0}. \label{eq:p2star}
\end{align}
\end{itemize}
\end{proposition}
\begin{IEEEproof}
See Appendix \ref{proof:ptilde}.
\end{IEEEproof}

Notice the solutions for $\alpha = 0$ and $\alpha = 1$ are not included in the proposition as they are trivial. For $\alpha = 0$, ZF-JT is applied in the whole frame. Then $\tilde{p} = 0$ and $p_i, i = 1, 2$ are obtained by solving (\ref{eq:power2gb}) and (\ref{eq:power3gb}). For $\alpha = 1$, the problem is feasible only when $A_{\bar{k}} = 0$, then $p_i = 0, i = 1, 2$ and $\tilde{p}$ can be obtained by solving (\ref{eq:power2gb}). According to Proposition \ref{prop:ptilde}, for $0 < \alpha < 1$, the power allocation problem (\ref{prob:poweralloc}) for the fixed $k$ and $\alpha$ with equality constraints (\ref{eq:power2gb}) and (\ref{eq:power3gb}) can be solved by calculating and comparing the values of $\tilde{p}_{\mathrm{min}}, \tilde{p}_{\mathrm{max}},$ and $\tilde{p}_0$. As they can be expressed in closed-form, the calculation is straightforward and simple. On the contrary, solving the original power allocation problem with inequality constraints (\ref{eq:power2p}) and (\ref{eq:power3p}) requires searching over the feasible set via iterations such as interior-point method \cite[Chap. 11]{boyd2004convex}.

\subsection{Optimization Over $\alpha$}
Besides the power allocation policy, we need to further determine optimal time ratio $\alpha$. As a matter of fact, the following theorem tells us that the optimization over $\alpha$ is also convex.

\begin{theorem} \label{prop:alphaconcave}
For a given $k$, define a function
\begin{equation}
F_k(\alpha) = \max_{\tilde{p}\!, p_1\!, p_2\!}  \alpha\!\log_2\!\Big(\!1\!+\!\frac{\tilde{p}|H_{\tilde{i}k}|^2}{\sigma^2_n}\!\Big) \!+ \! (\!1\!-\!\alpha\!)\sum_{i=1}^{2}\log_2\Big(\!1\!+\!\frac{p_{i}}{\sigma^2_n}\!\Big),  \label{func:alpha}
\end{equation}
where $0 \le \alpha \le 1$ and the maximization is constrained by (\ref{eq:power1p})-(\ref{eq:power4p}). $F_k(\alpha)$ is a concave function.
\end{theorem}
\begin{IEEEproof}
See Appendix \ref{proof:alphaconcave}.
\end{IEEEproof}

\begin{corollary}
The function $\bar{F}_k(\alpha) = F_k(\alpha)$, where the maximization is under constraints (\ref{eq:power1p}), (\ref{eq:power4p}), (\ref{eq:power2gb}), and (\ref{eq:power3gb}), is a concave function.
\end{corollary}
\begin{IEEEproof}
The proof simply follows the lines of Appendix \ref{proof:alphaconcave}.
\end{IEEEproof}

Since $\bar{F}_k(\alpha)$ is a concave function, the optimal $\alpha$ either satisfies $\bar{F}_k'(\alpha) = 0$ or takes the boundary values $\alpha_{\mathrm{min}}$ or $1$, where $\alpha_{\mathrm{min}} \le 1$ is presented in (\ref{eq:alphamin}) in Appendix \ref{proof:ptilde}. However, the closed-form solution for $\bar{F}_k'(\alpha) = 0$ is not easy to be obtained as the expression of $\bar{F}_k$ with respect to $\alpha$ is complex. Giving the condition that the value of $\bar{F}_k(\alpha)$ itself is easy to be computed, we can adopt the bi-section search algorithm and in each iteration check the monotonicity of $\bar{F}_k(\alpha)$ in a small neighborhood of $\alpha$. The bi-section search algorithm is detailed in Algorithm \ref{alg:bisection}.

\begin{algorithm}[th]
\caption{Bi-section search algorithm to find the maximum $\bar{F}_k(\alpha)$} \label{alg:bisection}
\begin{algorithmic}[1]

\STATE Initialize $\delta \alpha > 0, \uline{\alpha} = \alpha_{\mathrm{min}}, \bar{\alpha} = 1, I = 0$.

%\STATE Set $\hat{\alpha} = \frac{1}{2}(\uline{\alpha} + \bar{\alpha})$.

\WHILE {$I = 0$}

    \STATE Set $\hat{\alpha} = \frac{1}{2}(\uline{\alpha} + \bar{\alpha})$.

    \IF {$\bar{F}_k(\hat{\alpha}) \ge \bar{F}_k(\hat{\alpha} - \delta \alpha)$ and $\bar{F}_k(\hat{\alpha}) \ge \bar{F}_k(\hat{\alpha} + \delta \alpha)$}

        \STATE Set $I = 1$.

    \ELSE

        \IF {$\bar{F}_k(\hat{\alpha} - \delta \alpha) \le \bar{F}_k(\hat{\alpha}) \le \bar{F}_k(\hat{\alpha} + \delta \alpha)$}

            \STATE Set $\uline{\alpha} = \hat{\alpha}$.

        \ELSE

            \STATE Set $\bar{\alpha} = \hat{\alpha}$.

        \ENDIF

    \ENDIF

\ENDWHILE

\STATE The optimal solution is $\bar{F}_k(\hat{\alpha})$.

\end{algorithmic}
\end{algorithm}

In Algorithm \ref{alg:bisection}, $\delta \alpha$ should be carefully selected to balance the accuracy of the optimal solution $\hat{\alpha}$ and the convergence speed of the iteration. Before running the bi-section algorithm, we need to firstly check if the optimal is obtained at the boundary points. Altogether, the algorithm for calculating $\bar{g}(s, a)$ is summarized in Algorithm \ref{alg:perframeopt}.

\begin{algorithm}[th]
\caption{Per-stage Utility Calculation Algorithm} \label{alg:perframeopt}
\begin{algorithmic}[1]

%\STATE Find $P_N$ so that $\log_2(1+\gamma_NP_N) = B_N.$
%
%\STATE Set $\gamma_N' = \frac{1}{P_N + \frac{1}{\gamma_N}}$

\STATE Initialize $\bar{g}(s, a) = 0$ and $\delta \alpha > 0$.

\FORALL{$k = 1$ to $2$}

    \IF {${\cal P}_{k,\alpha_{\mathrm{min}}} \neq \emptyset$, and $\bar{F}_k(\alpha_{\mathrm{min}}) > \bar{F}_k(\alpha_{\mathrm{min}} + \delta \alpha)$}

        \STATE Update $\bar{g}(s, a) \leftarrow \max\{\bar{g}(s, a), \bar{F}_k(\alpha_{\mathrm{min}})\}$.

    \ELSIF {${\cal P}_{k,1} \neq \emptyset$, and $\bar{F}_k(1) > \bar{F}_k(1-\delta \alpha)$}

        \STATE Update $\bar{g}(s, a) \leftarrow \max\{\bar{g}(s, a), \bar{F}_k(1)\}$.

    \ELSE

        \STATE Run Algorithm \ref{alg:bisection}, and then update $\bar{g}(s, a) \leftarrow \max\{\bar{g}(s, a), \bar{F}_k(\hat{\alpha})\}$.

    \ENDIF

\ENDFOR

\end{algorithmic}
\end{algorithm}

\section{Approximate Dynamic Programming} \label{sec:ADP}
In this section, we adopt the approximate DP \cite[Chap.~6]{bertsekas2005dynamic} to solve the policy optimization problem and deal with the complexity issue due to the large number of system states. The basic idea of the approximate DP is to estimate the relative utility $h(s)$ via a set of parameters $\bm{c} = (c_1, c_2, \cdots, c_M)^T$ rather than to calculate the exact value. In this way, we only need to train the parameter vector $\bm{c}$ based on a small set of simulation samples. Specifically, we apply \emph{approximate policy iteration} algorithm as the convergence property can be guaranteed. Firstly, we briefly introduce the \emph{policy iteration} algorithm and its approximation version. Then we will implement the algorithm to solve our problem.

\subsection{Policy Iteration Algorithm}
The policy iteration algorithm includes two steps in each iteration: \emph{policy evaluation} and \emph{policy improvement}. It starts with any feasible stationary policy, and improves the objective step by step. Suppose in the $n$-th iteration, we have a stationary policy denoted by $\bm{a}^{(n)} = \{a^{(n)}(s)|s\in \mathcal{S}\}$. Based on this policy, we perform policy evaluation step, i.e., we solve the following linear equations
\begin{equation}
\lambda^{(n)} + h^{(n)}(s) = g(s, a^{(n)}(s)) + \sum_{s' \in \mathcal{S}}p_{s\rightarrow s' | a^{(n)}(s)}h^{(n)}(s') \label{eq:linear}
\end{equation}
for $\forall s \in \mathcal{S}$ to get the average cost $\lambda^{(n)}$ and the relative utility vector $\bm{h}^{(n)}$. Notice that the number of unknown parameters $(\lambda^{(n)}, \bm{h}^{(n)})$ is one more than the number of equations. Hence, more than one solutions exist, which are different with each other by a constant value for all $h^{(n)}(s)$. Without loss of generality, we can select a fixed state $s_0$ so that $h^{(n)}(s_0)=0$, then the solution for (\ref{eq:linear}) is unique.

The second step is to execute the policy improvement to find a stationary policy $\bm{a}^{(n+1)}$ which minimizes the right hand side of Bellman's equation
\begin{equation}
a^{(n+1)}(s) \!=\! \arg\!\max_{a\in \mathcal{A}(s)} \!\left[g(s, a) \!+\! \sum_{s' \in \mathcal{S}}\!p_{s\rightarrow s' | a}h^{(n)}(s')\right]. \label{eq:ukplus1}
\end{equation}

If $\bm{a^{(n+1)}} = \bm{a^{(n)}}$, the algorithm terminates, and the optimal policy is obtained $\bm{a^*} = \bm{a^{(n)}}$. Otherwise, repeat the procedure by replacing $\bm{a^{(n)}}$ with $\bm{a^{(n+1)}}$. It is proved that the policy \emph{does} improve the performance, i.e., $\lambda^{(n)} \le \lambda^{(n+1)}$ \cite[Prop.~4.4.2]{bertsekas2005dynamic}\footnote{For the average cost minimization problem discussed in Bertsekas's book, the direction of the inequality reverses.}, and the policy iteration algorithm terminates in finite number of iterations \cite[Prop.~4.4.1]{bertsekas2005dynamic}.

\subsection{Approximate Policy Evaluation}

For the policy evaluation step, the approximation DP tries to approximate the relative utility $h^{(n)}(s)$ by
\begin{equation}
\tilde{h}^{(n)}(s, \bm{c}^{(n)}) = \bm{\phi}(s)^T\bm{c}^{(n)}, \label{eq:hest}
\end{equation}
where $\bm{\phi}(s) = (\phi_1(s), \phi_2(s), \cdots, \phi_M(s))^T$ is an $M\times 1$ vector representing the features associated with state $s$, and $\bm{c}^{(n)} = (c_1^{(n)}, c_2^{(n)}, \cdots, c_M^{(n)})^T$ is an $M\times 1$ parameter vector. Instead of calculating all the relative utilities, we can train the parameter vector $\bm{c}^{(n)}$ using a relative small number of utility values and then estimate the others by (\ref{eq:hest}). Based on the estimated relative utility, the approximation of parameter vector for the next iteration is obtained by minimizing the least square error based on a weighted Euclidean norm, i.e.,
\begin{equation}
\bm{c}^{(n+1)} = \arg\min_{\bm{c} \in \mathcal{R}^M} ||\bm{\hat{h}}^{(n+1)} - \Phi\bm{c}||_{\bm{\xi}}^2,
\end{equation}
where $||\bm{J}||_{\bm{\xi}} = \sqrt{\sum_{s\in\mathcal{S}}\xi(s)J^2(s)}$ with a vector of positive weights $\xi(s), \forall s \in \mathcal{S}, \sum_s \xi(s) = 1$, $\mathcal{R}^M$ represents the $M$-dimensional real space, $\Phi$ is a matrix that has all the feature vectors $\phi(s)^T$, $\forall s \in \mathcal{S}$ as rows, and $\bm{\hat{h}}^{(n+1)} = F(\Phi\bm{c}^{(n)})$, where $F(\Phi\bm{c}^{(n)}) = (F(\bm{\phi}(s_1)^T\bm{c}^{(n)}), F(\bm{\phi}(s_2)^T\bm{c}^{(n)}), \cdots)^T$ and for each state $s$,
\begin{equation}
F(\bm{\phi}(s)^T\bm{c}^{(n)}) = g(s, a^{(n)}(s)) - \lambda^{(n)} + \sum_{s' \in \mathcal{S}}p_{s\rightarrow s' | a^{(n)}(s)}\phi(s')^T\bm{c}^{(n)}, \quad \forall s \in \mathcal{S}.
\end{equation}
%\begin{align}
%F(\bm{\phi}(s)^T\bm{c}^{(n)}) &= g(s, a^{(n)}(s)) - \lambda^{(n)} + \nonumber\\
%{}& \sum_{s' \in \mathcal{S}}p_{s\rightarrow s' | a^{(n)}(s)}\phi(s')^T\bm{c}^{(n)}, \quad \forall s \in \mathcal{S}.
%\end{align}
For simplicity, the mapping $F$ can be written in matrix form as in \cite[Sec 6.6]{bertsekas2005dynamic}, i.e., $F(\bm{h}) = \bm{g} - \lambda \bm{e} + \bm{P h}$, where $\lambda$ is the average utility, $\bm{P}$ is the transition probability matrix and $\bm{e}$ is the unit vector. Further more, the mapping $F$ can be replaced by a parameterized mapping $F^{(\beta)} = (1-\beta)\sum_{i = 0}^{+\infty}\beta^iF^{i+1}$, where $\beta \in [0, 1)$, and $F^{i+1}(\bm{h}) = F^{i}(F(\bm{h}))$. The algorithm is called \emph{least square policy evaluation with parameter $\beta$ (LSPE($\beta$))} \cite[Chap. 6]{bertsekas2005dynamic}. The benefit of introducing the parameter $\beta$ is as follows. On the one hand, a higher convergence rate and smaller error bound can be obtained by setting larger $\beta$. On the other hand, when simulation is applied for approximation, larger $\beta$ results in more pronounced simulation noise. Hence, tuning the parameter $\beta$ helps to balance these factors. If $\beta = 0$, the mapping reduces to $F$.

Actually, we do not need to calculate samples of ${\hat{h}}(s)$ to estimate $\bm{c}$. Instead, the calculation can be done by simulation. Specifically, we generate a long simulated trajectory $s_0, s_1, \cdots$ based on the given action $\bm{a}^{(n)}$, and update $\bm{c}$ for each simulation realization according to the least square error metric. The advantage of simulation is that we only need a simulated trajectory rather than the state transition probability for a given policy. In reality, it means that we can use the simulated samples or the historical samples to directly calculate the estimated relative utility, instead of firstly estimate the transition probability and then estimate the utility. In the simulation-based LSPE($\beta$) algorithm, $\bm{c}$ is updated iteratively according to each simulation sample. It can be expressed in matrix form \cite[Sec 6.6]{bertsekas2005dynamic} as for the $i$-th sample,
\begin{equation}
\bm{c}_{i+1} = \bm{c}_i + \bm{B}_i^{-1}(\bm{A}_i\bm{c}_i+\bm{b}_i), \label{eq:cLSPE}
\end{equation}
where
\begin{align}
\bm{A}_i &= \frac{i}{i+1}\bm{A}_{i-1} + \frac{1}{i+1}\bm{z}_i(\bm{\phi}(s_{i+1})^T-\bm{\phi}(s_{i})^T),\nonumber\\
\bm{B}_i &= \frac{i}{i+1}\bm{B}_{i-1} + \frac{1}{i+1}\bm{\phi}(s_{i})\bm{\phi}(s_{i})^T,\nonumber\\
\bm{b}_i &= \frac{i}{i+1}\bm{b}_{i-1} + \frac{1}{i+1}\bm{z}_i(g(s_i, a^{(n)}(s_i))-\lambda_i),\nonumber\\
\bm{z}_i &= \beta \bm{z}_{i-1} + \bm{\phi}(s_{i}), \nonumber\\
\lambda_i &= \frac{1}{i+1}\sum_{j=0}^ig(s_j, a^{(n)}(s_j)),\nonumber
\end{align}
for all $i \ge 0$ and the boundary values $\bm{A}_{-1} = 0, \bm{B}_{-1} = 0, \bm{b}_{-1} = 0, \bm{z}_{-1} = 0.$ Note that there are two iterations in the approximate DP. The outer iteration runs policy evaluation and policy improvement to update the policy, the inner iteration runs the LSPE($\beta$) algorithm to update the parameter vector $\bm{c}$. In the $n$-th policy evaluation, the policy $\bm{a}^{(n)}$ is viewed as an input to generate the simulation trajectory and calculate $\bm{c}_i$ according to (\ref{eq:cLSPE}) in the inner iteration. When the difference between $\bm{c}_{i+1}$ and $\bm{c}_i$ is small enough, the policy evaluation process terminates and we get $\bm{c}^{(n)} = \bm{c}_i$. Then the policy is updated using $\bm{c}^{(n)}$, i.e.,
\begin{equation}
a^{(n+1)}(s) \!=\! \arg\!\max_{a\in \mathcal{A}(s)} \!\left[g(s, a) \!+\! \sum_{s' \in \mathcal{S}}\!p_{s\rightarrow s' | a}\bm{\phi}(s')^T\bm{c}^{(n)}\right]. \nonumber
\end{equation}

Generally, the length of the simulation trajectory is small than the number system state. Hence, the computational complexity of policy evaluation step can be reduced, especially when the number of states is large. Notice that the policy improvement step still needs to go through all the states due to the existence of the maximization operation.

\subsection{Implementation Issues}
To get an efficient approximate DP algorithm, the features of each state $\bm{\phi}(s)$ needs to be carefully selected. In our problem, we consider the following features.

\begin{itemize}
 \item Energy-related features to indicate the influence of available energy on the utility. As the utility is represented in terms of data rate, the energy-related features are defined as $\log_2(1+\frac{B_k/T_f+E_k}{\sigma^2_n}), k = 1, 2.$
 \item Channel-related features to indicate the influence of channel gain. Similarly, they are defined as $\log_2(1+|H_{ik}|^2), i = 1, 2, k = 1, 2.$
 \item Cooperation features to indicate the influence of JT. As a MIMO system, the eigenvalues are the key indicator of the MIMO link performance. Hence, we define this type of feature as $\log_2(1+\rho_i), i = 1, 2,$ where $\rho_i, i = 1, 2$ are the eigenvalues of matrix $\mathbf{H}\mathbf{H}^H$.
 \item The 2nd-order features. As the actual data rate is calculated by the product of power and channel gain, we further consider the following features: $\log_2(1+\frac{(B_k/T_f+E_k)|H_{ik}|^2}{\sigma^2_n}), i = 1, 2, k = 1, 2$ and $\log_2(1+\frac{(B_k/T_f+E_k)\rho_k}{\sigma^2_n}), k = 1, 2$.
\end{itemize}

The second issue concerning the approximate DP is that as the estimated relative utility is calculated based on the simulation samples generated for a given policy. Thus, some states that are unlikely to occur under this policy are under-represented. As a result, the relative utility estimation of these states may be highly inaccurate, causing potentially serious errors in the policy improvement process. This problem is known as \emph{inadequate exploration} \cite[Sec. 6.2]{bertsekas2005dynamic} of the system dynamics. One possible way for guaranteeing adequate exploration of the state space is to frequently restart the simulation from a random state under a random policy. We call it as \emph{policy exploration}. We will show later in the next section the influence of policy exploration on the performance.

\section{Simulation Study} \label{sec:sim}
We study the performance of the proposed algorithms by simulations. We adopt the outdoor pico-cell physical channel model from 3GPP standard \cite{3GPP2010TR}. The pathloss is $\mathrm{PL} = 140.7 + 36.7\log_{10}d$ (dB), where the distance $d$ is measured in km. The distance between pico BSs is 100m. The shadowing fading follows log-normal distribution with variance 10dB. The small-scale fading follows Rayleigh distribution with zero mean and unit variance. The average SNR at the cell edge (50m to the pico BS) with transmit power 30dBm is set to 10dB.  We set the two users are placed in the cell edge of the two pico BSs depicted in Fig.~\ref{fig:system}. Hence, they experience the same large-scale fading. The BSs are equipped with energy harvesting devices (e.g.~solar panels). {The transmit power of pico BSs is around hundreds of mW, and we set the energy harvesting rate accordingly.}

\begin{figure}
\centering
\includegraphics[width=4.5in]{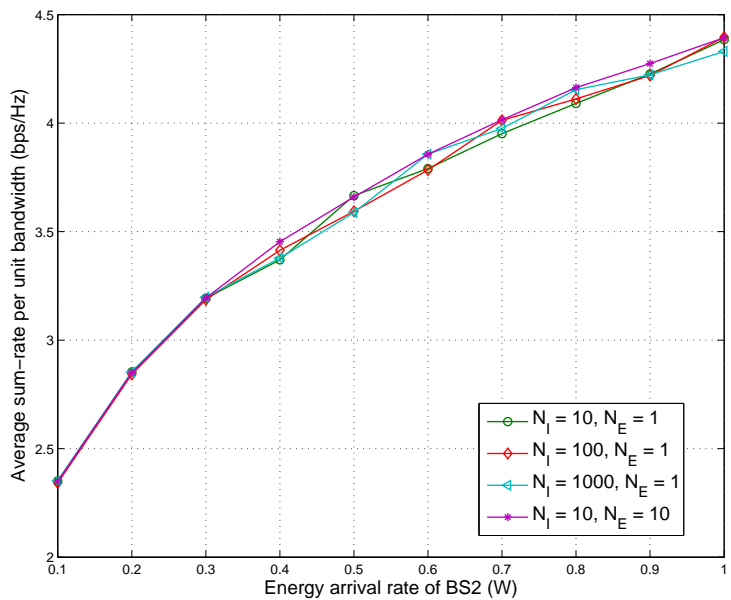}
\caption{The influence of number of iterations and number of policy explorations on the sum-rate performance of approximate DP. The energy arrival rate of BS1 is 0.1W.} \label{fig:LSPE}
\end{figure}

Firstly, we evaluate the influence of number of iterations in the approximate DP on the performance. We fix the energy arrival rate of BS1 as 0.1W and change that of BS2. Denote the number of iterations for policy improvement by $N_I$, and the number of policy explorations which restarts the policy iteration by $N_E$. We set different values of $N_I$ and $N_E$ to run the approximate DP algorithm and compare the achievable sum-rate. The result is shown in Fig.~\ref{fig:LSPE}. From this figure, we can see that if policy exploration is not considered, i.e., $N_E = 1$, the approximate DP reveals some random fluctuation. Solely increasing the number of policy iterations is not guaranteed to improve the performance. On the other hand, by increasing the number of policy explorations, the fluctuation can be efficiently reduced and the performance can be greatly improved, even with relatively small number of policy iterations. This validates the claim that the simulation-based policy iteration may be inaccurate, and it is quite important to adopt policy exploration in the approximate DP algorithm design.

\begin{figure}
\centering
\includegraphics[width=4.5in]{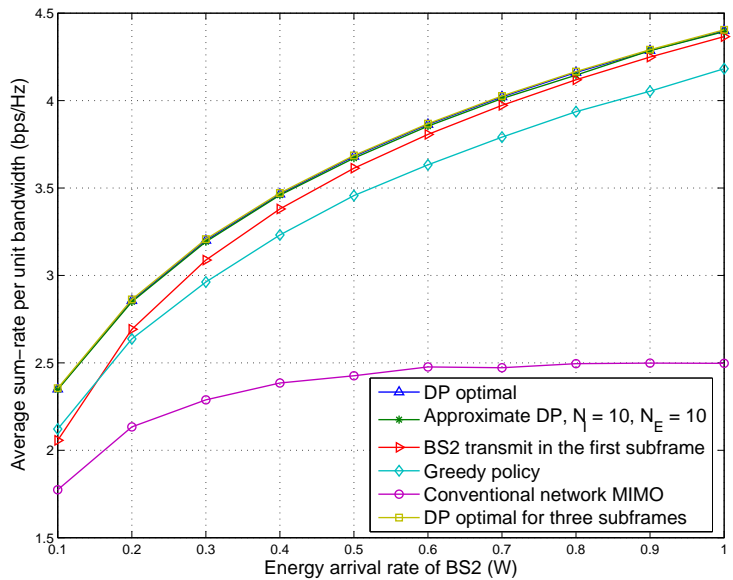}
\caption{Average sum-rate comparison of different algorithms. The energy arrival rate of BS1 is 0.1W.} \label{fig:compcmp}
\end{figure}

Then we show the performance of approximate DP compared with the optimal policy obtained via DP optimal algorithm. And the following baselines are also considered for comparison. In the conventional network MIMO, the whole frame applies ZF-JT without sub-frame spitting. In the greedy policy, we do not optimize the energy allocation among frames, but greedily use all the available energy for sum-rate maximization in each frame. Mathematically, we solve the problem (\ref{eq:frameprob}) under constraints (\ref{eq:power1p})-(\ref{eq:power5p}) with $A_k = \frac{B_k}{T_f}+E_k, k = 1,2.$ Hence, instead of finding the policy for each state before the system runs, we can get the online solution based on current system state. According to Theorem \ref{prop:convex} and Theorem \ref{prop:alphaconcave}, the problem can be solved by firstly applying bi-section search over $\alpha$ and then for each $\alpha$ calculating optimal power allocation via convex optimization. {Besides, we consider always selecting the BS with higher energy arrival rate to transmit in the single-BS transmission subframe. Finally, we also consider a more general fractional JT scheme that divides each frame into three subframes: Each BS transmits individually in the first and second subframe, and then they jointly transmit in the third subframe. We also solve the sum-rate maximization problem via DP.}

{By fixing the energy arrival rate of BS1 as 0.1W and changing that of BS2, the results are shown in Fig.~\ref{fig:compcmp}. It can be seen that the generalized fractional JT scheme with three subframes provides little performance gain compared with the scheme with two subframes, even with symmetric energy arrival rates. Intuitively, the fractional JT with three subframes may perform better in symmetric case. However, the performance depends not only on the energy arrival rates of two BSs, but also on the channel states. When the energy arrival rates are asymmetric, dividing each frame into two subframes and letting the BS with higher energy arrival rate to transmit in the first subframe is sufficient. When the energy arrival rates are symmetric, the channel states become the key factor. In fact, the case with asymmetric channel gains is analogous to the case with asymmetric energy profiles. Hence, letting the BS with higher channel gain to transmit in the first subframe is sufficient. The scheme with three subframes may be better in symmetric case, which is however of low probability as it requires the energy arrival rates and the channel states are jointly symmetric. In addition, the optimization for three subframes is much more complex than that for two subframes. Therefore, the fractional JT with two subframes is preferred.}

It can be also seen in Fig.~\ref{fig:compcmp} that the proposed approximate DP algorithm with $N_I = 10, N_E = 10$ performs very close to the optimal one. In addition, the greedy policy show a noticeable gap to the optimal policy, which illustrates the necessity of inter-frame energy allocation optimization. {Always choosing BS2 to transmit in the first subframe degrades the performance compared with the proposed algorithm, while the gap diminishes as the energy asymmetry becomes stronger. This is also due to the dependence of performance on both the energy profiles and the channel states. When the channel state of the BS with more energy is much worse than the other, it would be preferred to sleep to wait for a better channel.} Also, the proposed fractional JT algorithm dramatically outperforms the conventional network MIMO algorithm, especially when the asymmetry of energy arrival rate between two BSs becomes severe. {Notice that the performance gain is remarkable even for the symmetric case (energy arrival rate of BS2 is also 0.1W). As mentioned before, the gain comes from the asymmetry of channel states, which is analogous to the asymmetry of energy arrival rates.} With the increase of energy arrival rate in BS2, the sum-rate of conventional algorithm saturates to around 2.5bps/Hz. {The reason is that according to the power constraint (\ref{eq:powerctr}), the power constraint of BS2 associated with sufficiently large budget $P_{t,2}$ is usually satisfied with strict inequality. Then, increasing $P_{t,2}$ does not affect the optimization result. That is, the sum-rate does not increase as the higher energy arrival rate of BS2 does not contribute.} On the other hand, the sum-rate of the fractional JT increases in the speed of $\log$ function. It also shows the importance of applying fractional JT in energy harvesting system.

\begin{figure}
\centering
\includegraphics[width=4.5in]{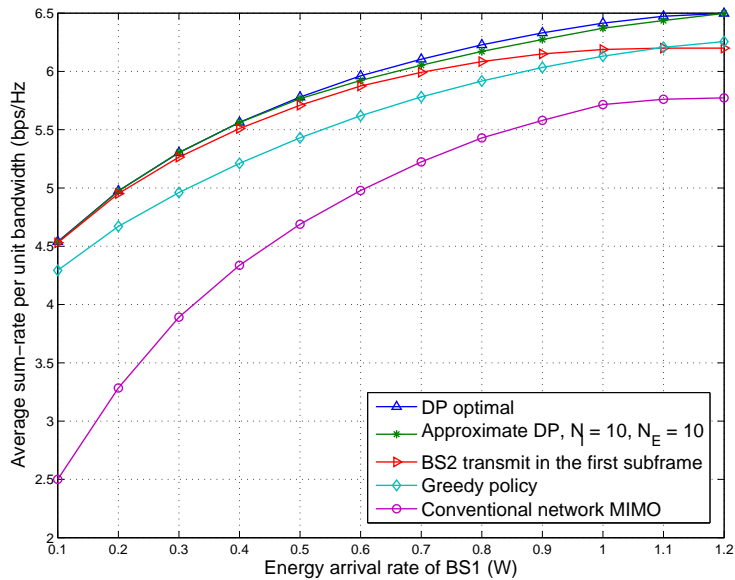}
\caption{Average sum-rate comparison of different algorithms. The energy arrival rate of BS2 is 1.2W.} \label{fig:compPmax}
\end{figure}

{We further simulate the case that the energy arrival rate is sufficient for transmission. We set the maximum transmit power per frame as 1.2W. The energy arrival rate of BS2 is equal to the maximum power per frame, and we vary the rate of BS1 to obtain the curves in Fig.~\ref{fig:compPmax}. It can be seen that the performance gain of the proposed fractional JT strategy compared with the conventional network MIMO decreases as the energy arrival rate of BS1 becomes closer to that of BS2. And all the curves tend to be flat when the maximum transmit power can be satisfied by energy harvesting. Besides, always choosing BS2 to transmit in the first subframe approaches optimal then the energy asymmetry is strong. But it performs even worth than the greedy policy in symmetric case when the maximum transmit power is achieved in both BSs.}

\begin{figure}
\centering
\includegraphics[width=4.5in]{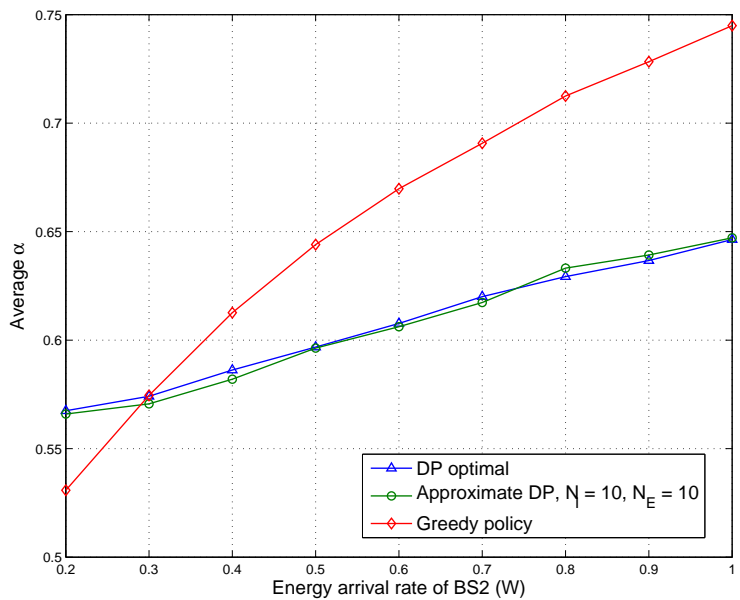}
\caption{Average time ratio $\alpha$ for single-transmission phase of different algorithms. The energy arrival rate of BS1 is 0.1W.} \label{fig:alpha}
\end{figure}

Fig.~\ref{fig:alpha} shows the average time ratio $\alpha$ for single-transmission phase versus the energy arrival rate of BS2. It can be seen that average $\alpha$ increases as the asymmetry of energy arrival rates increases. Furthermore, the average $\alpha$ of DP optimal algorithm increases at the lowest speed, and the approximate DP algorithm performs very close to it. The greedy policy can only increase the time ratio for single-transmission to better utilize the higher energy arrival rate, and hence $\alpha$ increases at a higher speed w.r.t. the increase of energy arrival rate of BS2. On the contrary, by averaging the available energy over the transmission frames in the DP optimal and approximate DP algorithms, relatively more time ratio can be used to apply network MIMO to enhance the sum-rate.

\begin{figure}
\centering
\includegraphics[width=4.5in]{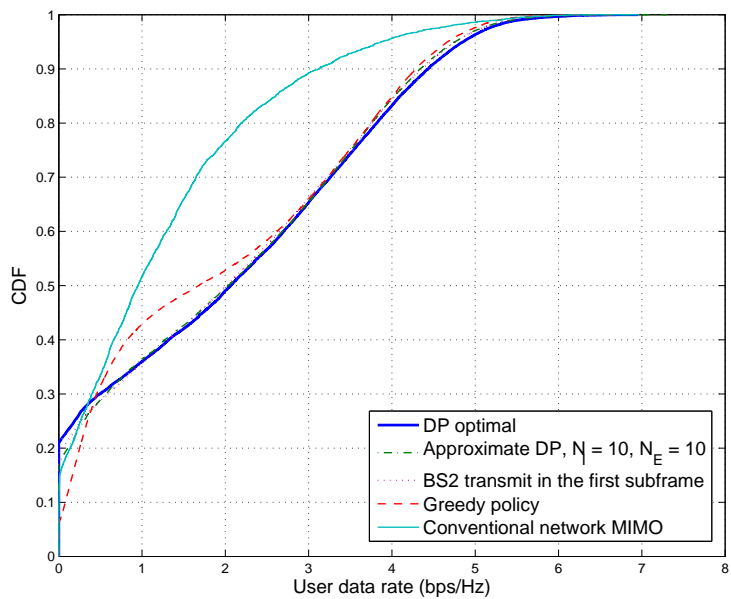}
\caption{Cumulative distribution function of user data rate with different algorithms. The energy arrival rate of BS1 is 0.1W, and that of BS2 is 0.8W.} \label{fig:CDF}
\end{figure}

Finally, the cumulative distribution function (CDF) of user data rate is depicted in Fig.~\ref{fig:CDF} with energy arrival rates of the two BSs as 0.1W and 0.8W, respectively. It shows that the proposed fractional JT algorithm greatly enhances the user data rate compared with the conventional network MIMO, and the proposed approximate DP algorithm achieves close-to-optimal performance. Since the energy arrival rate of BS2 is much larger than BS1, simply choosing BS2 to transmit in the first subframe also performs close to the optimal. Notice that the greedy policy reduces the percentage of zero data rate since it transmits with all the available energy in each frame, with the sacrifice of channel fading diversity for opportunistic inter-frame scheduling. As a result, the ratio of low data rate is much higher than the DP-based algorithms. For instance, about 43\% of users' data rate is lower than 1bps/Hz. With DP-based algorithms, the ratio reduces by about 8\%.

\section{Conclusion}\label{sec:concl}
In this paper, we have proposed a fractional JT scheme for BS cooperation that divides a transmission frame to firstly apply single-BS transmission and then adopt ZF-JT transmission to enhance the average sum-rate. The MDP-based problem is formulated and solved by firstly allocating energy among frames and then optimizing per-frame sum-rate. By analyzing the convexity of per-frame sum-rate optimization problem, and applying approximate DP algorithm, the computational complexity is greatly reduced. The proposed fractional JT scheme has been shown to achieve much higher sum-rate compared with the conventional ZF-JT only scheme. As the energy arrival asymmetry increases, the achievable rate of ZF-JT saturates (2.5bps/Hz in our settings), while the proposed scheme reveals a logarithmic increase. The proposed approximate DP algorithm can approach the DP optimal algorithm with sufficient number of policy explorations.

{ In this paper, fractional JT with two subframes is considered since we only consider the transmit power consumption. If the non-ideal circuit power is considered, more general frame structure is required to further save energy. Specifically, the BSs may turn to idle mode to reduce the circuit power consumption. This would be an interesting research direction for future work.}
%\section*{Acknowledgment}
%This work is sponsored in part by the National Science Foundation of China
%(NSFC) under grant No. 61201191, the National Basic Research Program of
%China (973 Program: 2012CB316001), the Creative Research Groups of NSFC
%under grant No. 61321061, and Hitachi R\&D Headquarter.

\appendices

\section{Proof of Proposition \ref{prop:greedy}} \label{proof:greedy}
For any given $\alpha$, the power allocation solution satisfies the Karush-Kuhn-Tucker (KKT) conditions \cite{boyd2004convex}. Define the Lagrangian function for any multipliers $\lambda\ge 0, \mu\ge 0, \eta\ge 0$ as
\begin{align}
\mathcal{L} = &-\Bigg(\alpha\!\log_2\!\Big(\!1\!+\!\frac{\tilde{p}|H_{\tilde{i}k}|^2} {\sigma^2_n}\!\Big) \!+ \! (\!1\!-\!\alpha\!)\sum_{i=1}^{2}\log_2\Big(\!1\!+\!\frac{p_{i}}{\sigma^2_n}\!\Big)\Bigg) +\lambda\Big(\alpha\tilde{p} - \frac{B_{k}}{T_f} - \alpha E_k\Big)\nonumber\\
&+\mu\Big((1-\alpha) \sum_{i=1}^2 |w_{ki}|^2 p_{i} + \alpha\tilde{p} - A_k\Big) +\eta\Big((1-\alpha)\sum_{i=1}^2 |w_{\bar{k}i}|^2 p_{i} - A_{\bar{k}}\Big) \label{eq:lagrangian}
\end{align}
%\begin{align}
%\mathcal{L} = &-\Bigg(\alpha\!\log_2\!\Big(\!1\!+\!\frac{\tilde{p}|H_{\tilde{i}k}|^2} {\sigma^2_n}\!\Big) \!+ \! (\!1\!-\!\alpha\!)\sum_{i=1}^{2}\log_2\Big(\!1\!+\!\frac{p_{i}}{\sigma^2_n}\!\Big)\Bigg) \nonumber\\
%&+\lambda\Big(\alpha\tilde{p} - \frac{B_{k}}{T_f} - \alpha E_k\Big) \nonumber\\
%&+\mu\Big((1-\alpha) \sum_{i=1}^2 |w_{ki}|^2 p_{i} + \alpha\tilde{p} - A_k\Big) \nonumber\\
%&+\eta\Big((1-\alpha)\sum_{i=1}^2 |w_{\bar{k}i}|^2 p_{i} - A_{\bar{k}}\Big) \label{eq:lagrangian}
%\end{align}
with additional complementary slackness conditions
\begin{align}
\lambda\Big(\alpha\tilde{p} - \frac{B_{k}}{T_f} - \alpha E_k\Big) &= 0, \nonumber\\
\mu\Big((1-\alpha) \sum_{i=1}^2 |w_{ki}|^2 p_{i} + \alpha\tilde{p} - A_k\Big) &=0,\nonumber\\
\eta\Big((1-\alpha)\sum_{i=1}^2 |w_{\bar{k}i}|^2 p_{i} - A_{\bar{k}}\Big) &= 0.\nonumber
\end{align}

Here, we ignore the non-negative power constraints in the above formulation to simplify the expression. It can be directly added to the result. We apply the KKT optimality conditions to the Lagrangian function (\ref{eq:lagrangian}). By setting $\partial\mathcal{L}/\partial \tilde{p} = \partial\mathcal{L}/\partial p_{i} =0$, we obtain
\begin{eqnarray}
\tilde{p}^* & = & \left[\frac{1}{\lambda+\mu}-\frac{\sigma^2_n}{|H_{\tilde{i}k}|^2}\right]^+, \label{opt:ptilde}\\
p_i^* & = & \left[\frac{1}{\mu|w_{ki}|^2+|w_{\bar{k}i}|^2\eta}-\sigma^2_n\right]^+, i = 1, 2.
\label{opt:pj}
\end{eqnarray}

Notice that to guarantee the validity of (\ref{opt:pj}), either $\mu$ or $\eta$ should be non-zero, which means that at least one of (\ref{eq:power2p}) and (\ref{eq:power3p}) is satisfied with equality.

\section{Proof of Lemma \ref{lemma:hincre}} \label{proof:hincre}
Since $h^*(s) = \lim\limits_{n\rightarrow+\infty}h^{(n)}(s)$, we prove the monotonicity property by induction. In addiction, we only need to prove the monotonicity for $B_1$. The proof for $B_2$ follows the same procedure.

Obviously, it is true for $n=0$ as $h^{(0)}(s) = 0, \forall s \in \mathcal{S}$. Assume that $h^{(n)}(B_1, B_2, \mathbf{H})$ is nondecreasing w.r.t $B_1$, and the optimal action for state $s = (B_1, B_2, \mathbf{H})$ is $a^* = (A_1^*, A_2^*)$, i.e.,
\begin{align}
\max_{a\in \mathcal{A}(s)}\left[ g(s, a) + \tau\sum_{ \mathbf{H}'} \mathrm{Pr} (\mathbf{H}'|\mathbf{H})h^{(n)}(s')\right]
=  g(s, A_1^*, A_2^*) + \tau\sum_{ \mathbf{H}'} \mathrm{Pr} (\mathbf{H}'|\mathbf{H})h^{(n)}(B_1', B_2', \mathbf{H}'). \nonumber
\end{align}
%\begin{align}
%&\max_{a\in \mathcal{A}(s)}\left[ g(s, a) + \tau\sum_{ \mathbf{H}'} \mathrm{Pr} (\mathbf{H}'|\mathbf{H})h^{(n)}(s')\right]\nonumber\\
%= & g(s, A_1^*, A_2^*) + \tau\sum_{ \mathbf{H}'} \mathrm{Pr} (\mathbf{H}'|\mathbf{H})h^{(n)}(B_1', B_2', \mathbf{H}'). \nonumber
%\end{align}
Then consider the state $s'' = (B_1+\delta B, B_2, \mathbf{H})$, where $\delta B > 0$. We have
\begin{align}
&h^{(n+1)}(s'')\nonumber\\
= &(1-\tau)h^{(n)}(s'') + \max_{a\in \mathcal{A}(s'')}\left[ g(s'', a) + \tau\sum_{ \mathbf{H}'} \mathrm{Pr} (\mathbf{H}'|\mathbf{H})h^{(n)}(s')\right] - \Lambda^{(n+1)}(s_0) \nonumber\\
\buildrel (a) \over \ge &(1-\tau)h^{(n)}(s'') + g(s'', A_1^*, A_2^*) + \tau\sum_{ \mathbf{H}'} \mathrm{Pr} (\mathbf{H}'|\mathbf{H})h^{(n)}(B_1'+\delta B, B_2', \mathbf{H}') - \Lambda^{(n+1)}(s_0) \nonumber\\
\buildrel (b) \over \ge &(1-\tau)h^{(n)}(s) + g(s, A_1^*, A_2^*) + \tau\sum_{ \mathbf{H}'} \mathrm{Pr} (\mathbf{H}'|\mathbf{H})h^{(n)}(B_1', B_2', \mathbf{H}') - \Lambda^{(n+1)}(s_0)
= h^{(n+1)}(s), \nonumber
\end{align}
%\begin{align}
%&h^{(n+1)}(s'')\nonumber\\
%= &(1-\tau)h^{(n)}(s'') + \max_{a\in \mathcal{A}(s'')}\left[ g(s'', a) +\right. \nonumber\\
%{}&\qquad\qquad\qquad \left. \tau\sum_{ \mathbf{H}'} \mathrm{Pr} (\mathbf{H}'|\mathbf{H})h^{(n)}(s')\right] - \Lambda^{(n+1)}(s_0) \nonumber\\
%\buildrel (a) \over \ge &(1-\tau)h^{(n)}(s'') + g(s'', A_1^*, A_2^*) + \nonumber\\
%{}&\; \tau\sum_{ \mathbf{H}'} \mathrm{Pr} (\mathbf{H}'|\mathbf{H})h^{(n)}(B_1'+\delta B, B_2', \mathbf{H}') - \Lambda^{(n+1)}(s_0) \nonumber\\
%\buildrel (b) \over \ge &(1-\tau)h^{(n)}(s) + g(s, A_1^*, A_2^*) + \nonumber\\
%{}&\qquad\quad \tau\sum_{ \mathbf{H}'} \mathrm{Pr} (\mathbf{H}'|\mathbf{H})h^{(n)}(B_1', B_2', \mathbf{H}') - \Lambda^{(n+1)}(s_0) \nonumber\\
%= &h^{(n+1)}(s), \nonumber
%\end{align}
where the inequality (a) holds as the action $(A_1^*, A_2^*) \in \mathcal{A}(s'')$, and (b) holds due to the following two reasons. Firstly, $g(s'', A_1^*, A_2^*) \ge g(s, A_1^*, A_2^*)$ as the constraint (\ref{eq:power1p}) for the latter is not looser than the former. Secondly, $h^{(n)}(B_1'+\delta B, B_2', \mathbf{H}') \ge h^{(n)}(B_1', B_2', \mathbf{H}')$ due to the monotonicity of $h^{(n)}(B_1, B_2, \mathbf{H})$ w.r.t. $B_1$. As a result, we prove that $h^{(n+1)}(B_1, B_2, \mathbf{H})$ is also nondecreasing w.r.t. $B_1$.

In summary, $h^{(n)}(B_1, B_2, \mathbf{H})$ is nondecreasing w.r.t. $B_1$ for all $n = 0, 1, 2, \cdots$. Hence, we also have that $h^*(B_1, B_2, \mathbf{H})$ is nondecreasing w.r.t. $B_1$. The same holds for $B_2$.

\section{Proof of Theorem \ref{prop:gbar}} \label{proof:gbar}
Regarding the per-stage utility $\bar{g}$, the Bellman's equation also holds for a scalar $\bar{\Lambda}^*$ and some vector $\bm{\bar{h}}^* = \{\bar{h}^*(s)|s \in \mathcal{S}\}$, and the value iteration algorithm works in the same way. Hence, we only need to prove by induction that $\Lambda^{(n)}(s_0) = \bar{\Lambda}^{(n)}(s_0)$ and $h^{(n)}(s) = \bar{h}^{(n)}(s)$.

We initialize that $\Lambda^{(0)}(s_0) = \bar{\Lambda}^{(0)}(s_0) = 0$ and $h^{(0)}(s) = \bar{h}^{(0)}(s) = 0, \forall s \in \mathcal{S}$. Suppose that $\Lambda^{(n)}(s_0) = \bar{\Lambda}^{(n)}(s_0), h^{(n)}(s) = \bar{h}^{(n)}(s), \forall s \in \mathcal{S}$. For the $(n+1)$-th iteration and $\forall s = (B_1, B_2, \mathbf{H}), a = (A_1, A_2)$, we have
\begin{align}
\bar{g}(s, a) + \tau\sum_{ \mathbf{H}'} \mathrm{Pr} (\mathbf{H}'|\mathbf{H})h^{(n)}(B_1', B_2', \mathbf{H}')
\buildrel (c) \over \le & {g}(s, a) + \tau\sum_{ \mathbf{H}'} \mathrm{Pr} (\mathbf{H}'|\mathbf{H})h^{(n)}(B_1', B_2', \mathbf{H}') \nonumber\\
\buildrel (d) \over \le & {g}(s, a) + \tau\sum_{ \mathbf{H}'} \mathrm{Pr} (\mathbf{H}'|\mathbf{H})h^{(n)}(B_1'', B_2'', \mathbf{H}')\nonumber
\end{align}
%\begin{align}
%&\bar{g}(s, a) + \tau\sum_{ \mathbf{H}'} \mathrm{Pr} (\mathbf{H}'|\mathbf{H})h^{(n)}(B_1', B_2', \mathbf{H}') \nonumber\\
%\buildrel (c) \over \le & {g}(s, a) + \tau\sum_{ \mathbf{H}'} \mathrm{Pr} (\mathbf{H}'|\mathbf{H})h^{(n)}(B_1', B_2', \mathbf{H}') \nonumber\\
%\buildrel (d) \over \le & {g}(s, a) + \tau\sum_{ \mathbf{H}'} \mathrm{Pr} (\mathbf{H}'|\mathbf{H})h^{(n)}(B_1'', B_2'', \mathbf{H}')\nonumber
%\end{align}
where $B_k' = B_k + T_fE_k - A_k, \forall k = 1, 2,$ while $B_k'', k = 1, 2$ are calculated via (\ref{eq:battery1}) and (\ref{eq:battery2}), respectively. Hence we have $B_k'' \ge B_k', \forall k = 1, 2$. Inequality (c) holds as the maximization of $g$ has larger feasible region than that of $\bar{g}$, while (d) holds due to the monotonicity of the relative utility $h(s)$. As a result, we have
\begin{align}
\max_{a \in \mathcal{A}(s)}\left[\bar{g}(s, a) + \tau\sum_{ \mathbf{H}'} \mathrm{Pr} (\mathbf{H}'|\mathbf{H})\bar{h}^{(n)}(s')\right] \le \max_{a \in \mathcal{A}(s)}\left[g(s, a) + \tau\sum_{ \mathbf{H}'} \mathrm{Pr} (\mathbf{H}'|\mathbf{H})h^{(n)}(s')\right] \label{proof:le}
\end{align}
%\begin{align}
%{}&\max_{a \in \mathcal{A}(s)}\left[\bar{g}(s, a) + \tau\sum_{ \mathbf{H}'} \mathrm{Pr} (\mathbf{H}'|\mathbf{H})\bar{h}^{(n)}(s')\right] \nonumber\\
%{}&\le \max_{a \in \mathcal{A}(s)}\left[g(s, a) + \tau\sum_{ \mathbf{H}'} \mathrm{Pr} (\mathbf{H}'|\mathbf{H})h^{(n)}(s')\right] \label{proof:le}
%\end{align}
On the other hand, there exists an action $(A_1^*, A_2^*)$ such that
\begin{align}
\max_{a\in \mathcal{A}(s)}\left[ g(s, a) + \tau\sum_{ \mathbf{H}'} \mathrm{Pr} (\mathbf{H}'|\mathbf{H})h^{(n)}(s')\right]
= & g(s, A_1^*, A_2^*) + \tau\sum_{ \mathbf{H}'} \mathrm{Pr} (\mathbf{H}'|\mathbf{H})h^{(n)} (B_1^*, B_2^*, \mathbf{H}'),\nonumber\\
\buildrel (e) \over = & \bar{g}(s, A_1^*, A_2^*) + \tau\sum_{ \mathbf{H}'} \mathrm{Pr} (\mathbf{H}'|\mathbf{H})\bar{h}^{(n)} (B_1^*, B_2^*, \mathbf{H}'),\nonumber\\
\buildrel (f) \over \le & \max_{a \in \mathcal{A}(s)}\left[\bar{g}(s, a) + \tau\sum_{ \mathbf{H}'} \mathrm{Pr} (\mathbf{H}'|\mathbf{H})\bar{h}^{(n)}(s')\right],\label{proof:ge}
\end{align}
%\begin{align}
%&\max_{a\in \mathcal{A}(s)}\left[ g(s, a) + \tau\sum_{ \mathbf{H}'} \mathrm{Pr} (\mathbf{H}'|\mathbf{H})h^{(n)}(s')\right]\nonumber\\
%= & g(s, A_1^*, A_2^*) + \tau\sum_{ \mathbf{H}'} \mathrm{Pr} (\mathbf{H}'|\mathbf{H})h^{(n)} (B_1^*, B_2^*, \mathbf{H}'),\nonumber\\
%\buildrel (e) \over = & \bar{g}(s, A_1^*, A_2^*) + \tau\sum_{ \mathbf{H}'} \mathrm{Pr} (\mathbf{H}'|\mathbf{H})\bar{h}^{(n)} (B_1^*, B_2^*, \mathbf{H}'),\nonumber\\
%\buildrel (f) \over \le & \max_{a \in \mathcal{A}(s)}\left[\bar{g}(s, a) + \tau\sum_{ \mathbf{H}'} \mathrm{Pr} (\mathbf{H}'|\mathbf{H})\bar{h}^{(n)}(s')\right],\label{proof:ge}
%\end{align}
where $B_k^* = B_k + T_fE_k - A_k^*, \forall k = 1, 2$, and hence, equality (e) holds. Inequality (f) holds as $(A_1^*, A_2^*) \in \mathcal{A}(s)$. It can be seen by (\ref{proof:le}), (\ref{proof:ge}) jointly with (\ref{eq:lambdanplus1}) and (\ref{eq:hnplus1}) that $\Lambda^{(n+1)}(s_0) = \bar{\Lambda}^{(n+1)}(s_0)$ and $h^{(n+1)}(s) = \bar{h}^{(n+1)}(s)$.

In summary, we have $\Lambda^{(n)}(s_0) = \bar{\Lambda}^{(n)}(s_0), h^{(n)}(s) = \bar{h}^{(n)}(s)$ for all $n = 0, 1, 2, \cdots$. Hence, we have $\Lambda^* = \max \; \lim_{N\rightarrow \infty}\mathbb{E}_{\mathbf{H}}\!\left[\!\frac{1}{N}\sum_{t=1}^N \bar{g}(s_t, a_t(s_t))\!\right] = \bar{\Lambda}^*$.

\section{Proof of Proposition \ref{prop:ptilde}} \label{proof:ptilde}
According to the equality constraints (\ref{eq:power2gb}) and (\ref{eq:power3gb}), $p_i, i = 1, 2$ can be represented as functions of $\tilde{p}$, i.e., $p_1 = \frac{C_{1} - \alpha|w_{\bar{k}2}|^2\tilde{p}}{C_0}, p_2 = \frac{\alpha|w_{\bar{k}1}|^2\tilde{p} - C_{2}}{C_0}$, where $C_0, C_{1}, C_{2}$ are presented in the proposition.
As the elements of $\mathbf{H}$ are i.i.d., we have $C_0 \neq 0$. Hence, the per-stage sum rate function can be written as a function of $\tilde{p}$:
\begin{align}
f_{k,\alpha}(\tilde{p}) = \alpha\!\log_2\!\Big(\!1\!+\!\frac{\tilde{p}|H_{\tilde{i}k}|^2} {\sigma^2_n}\!\Big) \!+ \! (\!1\!-\!\alpha\!)\left[\log_2\Big(\!1\!+\!\frac{C_{1} - \alpha|w_{\bar{k}2}|^2\tilde{p}}{\sigma^2_n C_0}\!\Big) + \log_2\Big(\!1\!+\!\frac{ \alpha|w_{\bar{k}1}|^2\tilde{p} - C_{2}}{\sigma^2_n C_0}\!\Big)\right]. \nonumber
\end{align}
%\begin{align}
%f_{k,\alpha}(\tilde{p}) = &\alpha\!\log_2\!\Big(\!1\!+\!\frac{\tilde{p}|H_{\tilde{i}k}|^2} {\sigma^2_n}\!\Big) \!+ \nonumber\\
%{}&(\!1\!-\!\alpha\!)\left[\log_2\Big(\!1\!+\!\frac{C_{1} - \alpha|w_{\bar{k}2}|^2\tilde{p}}{\sigma^2_n C_0}\!\Big) + \right. \nonumber\\
%{}&\left. \log_2\Big(\!1\!+\!\frac{ \alpha|w_{\bar{k}1}|^2\tilde{p} - C_{2}}{\sigma^2_n C_0}\!\Big)\right].
%\end{align}
The constraints can be written as the feasible set of $\tilde{p}$. Without loss of generality, we assume $C_0 > 0$. The feasible set for $C_0 < 0$ can be derived in the similar way. With the  non-negative constraints $p_i \ge 0, i = 1, 2$, we have $\frac{C_{2}}{\alpha|w_{\bar{k}1}|^2} \le \tilde{p} \le \frac{C_{1}}{\alpha|w_{\bar{k}2}|^2}$. Jointly with (\ref{eq:power1p}) and $\tilde{p} \ge 0$, the feasible set can be expressed as ${\cal P}_{k,\alpha} = \left\{\tilde{p} \Big| \tilde{p}_{\mathrm{min}} \le \tilde{p} \le \right. \left.\tilde{p}_{\mathrm{max}} \right\}$, where $\tilde{p}_{\mathrm{min}}$ and $\tilde{p}_{\mathrm{max}}$ are expressed as (\ref{ptilde:min}) and (\ref{ptilde:max}), respectively. To guarantee that ${\cal P}_{k,\alpha} \neq \emptyset$, we have $\tilde{p}_{\mathrm{min}} \le \tilde{p}_{\mathrm{max}}$, which results in $\alpha \ge \frac{1}{E_k}\big( \frac{C_{2}}{|w_{\bar{k}1}|^2} - \frac{B_k}{T_f}\big)$. We set
\begin{equation}
\alpha_{\mathrm{min}} = \max\left\{0, \frac{1}{E_k}\Big( \frac{C_{2}}{|w_{\bar{k}1}|^2} - \frac{B_k}{T_f}\Big)\right\}.  \label{eq:alphamin}
\end{equation}
Hence, there are two cases so that ${\cal P}_{k,\alpha} = \emptyset$. The first is $\alpha_{\mathrm{min}} > 1$, and the second is that $0 < \alpha_{\mathrm{min}} \le 1$ and $0 \le \alpha < \alpha_{\mathrm{min}}$. Otherwise, the per-frame optimization problem can be reformulated as
\begin{equation}
\max_{\tilde{p} \in {\cal P}_{k,\alpha}} f_{k,\alpha}(\tilde{p}),  \label{eq:maxptilde}
\end{equation}
whose convexity still holds according to the following lemma.

\begin{lemma}
The problem (\ref{eq:maxptilde}) is a convex optimization problem.
\end{lemma}
\begin{IEEEproof}
As the $\log$ function is concave and the functions inside the $\log$ operation are linear function of $\tilde{p}$, the composition of a linear function with a concave function is still concave. Hence, $f_{k,\alpha}(\tilde{p})$ is a concave function. On the other hand, the feasible set ${\cal P}_{k,\alpha}$ is convex. Therefore, the considered problem is a convex optimization problem.
\end{IEEEproof}

Due to the concavity of the function $f_{k,\alpha}(\tilde{p})$, the optimal solution can be found by solving $f_{k,\alpha}'(\tilde{p}) = 0$, which is expressed as (\ref{eq:quadratic}). It can be transformed into a quadratic equation, and hence, the nonnegative root can be easily solved. Denote the solution for $f_{k,\alpha}'(\tilde{p}) = 0$ by $\tilde{p}_0$. Then according to the concavity of the function $f_{k,\alpha}$, the optimal solution for the problem $\max\limits_{\tilde{p} \in {\cal P}_{k,\alpha}} f_{k,\alpha}(\tilde{p})$ is either $\tilde{p}_0$ or the boundary points of the feasible set ${\cal P}_{k,\alpha}$ depending on whether $\tilde{p}_0 \in {\cal P}_{k,\alpha}$ or not.

\section{Proof of Theorem \ref{prop:alphaconcave}} \label{proof:alphaconcave}
For any $\alpha^{(1)}, \alpha^{(2)} \in [0, 1]$, we assume that
\begin{equation}
F_k(\alpha^{(j)}) = \alpha^{(j)}\!\log_2\!\Big(\!1\!+\!\frac{\tilde{p}^{(j)}|H_{\tilde{i}k}|^2}{\sigma^2_n}\!\Big) \!+ \! (\!1\!-\!\alpha^{(j)}\!)\sum_{i=1}^{2}\!\log_2\!\Big(\!1\!+\!\frac{p_{i}^{(j)}}{\sigma^2_n}\!\Big), \nonumber
\end{equation}
for $j = 1, 2$, i.e., $\tilde{p}^{(j)}, p_{i}^{(j)}, i = 1, 2$ achieve the maximum sum-rate. For any $0 < \gamma < 1$, we have
\begin{align}
{}\gamma F_k(\alpha^{(1)}) + (1-\gamma) F_k(\alpha^{(2)})
\le\alpha'\!\log_2\!\Big(\!1\!+\!\frac{\tilde{p}'|H_{\tilde{i}k}|^2}{\sigma^2_n}\!\Big) \!+ \! (\!1\!-\!\alpha'\!)\sum_{i=1}^{2}\log_2\Big(\!1\!+\!\frac{p_{i}'}{\sigma^2_n}\!\Big) \label{eq:concave1}
\end{align}
%\begin{align}
%{}&\gamma F_k(\alpha^{(1)}) + (1-\gamma) F_k(\alpha^{(2)}) \nonumber\\
%\le&\alpha'\!\log_2\!\Big(\!1\!+\!\frac{\tilde{p}'|H_{\tilde{i}k}|^2}{\sigma^2_n}\!\Big) \!+ \! (\!1\!-\!\alpha'\!)\sum_{i=1}^{2}\log_2\Big(\!1\!+\!\frac{p_{i}'}{\sigma^2_n}\!\Big) \label{eq:concave1}
%\end{align}
where
\begin{align}
\alpha' =& \gamma\alpha^{(1)} + (1-\gamma)\alpha^{(1)}, \label{eq:alphap}\\
\tilde{p}' =& \frac{\gamma\alpha^{(1)}}{{\alpha'}}\tilde{p}^{(1)} +\frac{(1 - \gamma) \alpha^{(2)}}{{\alpha'}}\tilde{p}^{(2)}, \nonumber\\
p_i' = & \frac{\gamma(1-\alpha^{(1)})}{{1-\alpha'}}{p}_i^{(1)} +\frac{(1 - \gamma) (1-\alpha^{(2)})}{{1-\alpha'}}{p}_i^{(2)}, \quad i = 1, 2, \nonumber
\end{align}
and the inequality in (\ref{eq:concave1}) is due to the concavity of $\log$ function. In addition,
\begin{align}
\alpha'\tilde{p}' =& {\gamma\alpha^{(1)}}\tilde{p}^{(1)} +{(1 - \gamma) \alpha^{(2)}}\tilde{p}^{(2)} \nonumber\\
{}\le& \gamma\left( \frac{B_k}{T_f}+\alpha^{(1)}E_k \right) +(1 - \gamma) \left( \frac{B_k}{T_f}+\alpha^{(2)}E_k \right)
= \frac{B_k}{T_f} + \alpha'E_k, \nonumber
\end{align}
%\begin{align}
%\alpha'\tilde{p}' =& {\gamma\alpha^{(1)}}\tilde{p}^{(1)} +{(1 - \gamma) \alpha^{(2)}}\tilde{p}^{(2)} \nonumber\\
%{}\le& \gamma\left( \frac{B_k}{T_f}+\alpha^{(1)}E_k \right) +(1 - \gamma) \left( \frac{B_k}{T_f}+\alpha^{(2)}E_k \right) \nonumber\\
%=& \frac{B_k}{T_f} + \alpha'E_k, \nonumber
%\end{align}
i.e., $\tilde{p}'$ satisfies the constraint (\ref{eq:power1p}). Similarly, $\tilde{p}'$ and $p_i', i = 1, 2$ also satisfy the constraints (\ref{eq:power2p}) and (\ref{eq:power3p}). Hence, $\tilde{p}', p_i', i = 1, 2$ is a feasible power allocation solution. As $F_k(\alpha)$ is maximal over all power allocation policies, we have
\begin{equation}
\alpha'\!\log_2\!\Big(\!1\!+\!\frac{\tilde{p}'|H_{\tilde{i}k}|^2}{\sigma^2_n}\!\Big) \!+ \! (\!1\!-\!\alpha'\!)\sum_{i=1}^{2}\log_2\Big(\!1\!+\!\frac{p_{i}'}{\sigma^2_n}\!\Big) \le F_k(\alpha'). \label{eq:concave2}
\end{equation}
Combining (\ref{eq:concave1}), (\ref{eq:alphap}) and (\ref{eq:concave2}), we have
\begin{equation}
\gamma F_k(\alpha^{(1)}) + (1-\gamma) F_k(\alpha^{(2)}) \le F_k(\gamma\alpha^{(1)} + (1-\gamma)\alpha^{(1)}). \nonumber
\end{equation}
As a consequence, $F_k$ is a concave function.



%\bibliographystyle{IEEEtran}
%\begin{thebibliography}{1}
\bibliographystyle{IEEEtran}
\bibliography{ref}

%\end{thebibliography}

\end{document}